\documentclass[journal]{IEEEtran}
\usepackage{cite}
\ifCLASSINFOpdf
  \usepackage[pdftex]{graphicx}
\else
  \usepackage[dvips]{graphicx}
\fi
\usepackage{amsmath}
\usepackage{amsthm}
\usepackage{hyperref}
\usepackage{graphics}
\usepackage{algorithmic}
\usepackage{url}
\usepackage{array}
\usepackage{subfigure}
\usepackage{hhline}
\usepackage{url}
\usepackage{multirow}
\usepackage{wrapfig}
\usepackage{hyperref}
\usepackage{color}
\usepackage{bm}
\usepackage{embrac}
\usepackage{textcomp}
\usepackage{tabularx}

\mathchardef\mhyphen="2D
\newcounter{lfigcounter}

\begin{document}
%
\title{Medical Image Segmentation Using Deep Learning: A Survey}
%
%
%

\author{Risheng Wang, Tao Lei, Ruixia Cui, Bingtao Zhang, Hongying Meng and  Asoke K. Nandi
\thanks{R. Wang and T. Lei are with the School of Electronic Information and Artificial Intelligence and the Shaanxi Joint Laboratory of Artificial Intelligence, Shaanxi University of Science and Technology, Xi’an 710021, China.}
\thanks{R. Cui is with the 'Laboratory of Hepatobiliary Surgery, First Affiliated Hospital' and 'National Engineering Laboratory of Big Data Algorithm and Analysis Technology Research'(Xi’an Jiaotong University), Xi’an, 710049, China.}
\thanks{B. Zhang is with the School of Electronic and Information Engineering, Lanzhou Jiaotong University, Lanzhou 730070, China.}
\thanks{H. Meng is with the Department of Electronic and Electrical Engineering, Brunel University London, Uxbridge UB8 3PH, U.K.}
\thanks{A. K. Nandi is with the Department of Electronic and Electrical Engineering, Brunel University London, Uxbridge UB8 3PH, U.K.}
\thanks{(Corresponding author: Tao Lei) (E-mail: leitao@sust.edu.cn)}
}


%



\maketitle

\begin{abstract}
Deep learning has been widely used for medical image segmentation and a large number of papers has been presented recording the success of deep learning in the field. In this paper, we present a comprehensive thematic survey on medical image segmentation using deep learning techniques. This paper makes two original contributions. Firstly, compared to traditional surveys that directly divide literatures of deep learning on medical image segmentation into many groups and introduce literatures in detail for each group, we classify currently popular literatures according to a multi-level structure from coarse to fine. Secondly, this paper focuses on supervised and weakly supervised learning approaches, without including unsupervised approaches since they have been introduced in many old surveys and they are not popular currently. For supervised learning approaches, we analyze literatures in three aspects: the selection of backbone networks, the design of network blocks, and the improvement of loss functions. For weakly supervised learning approaches, we investigate literature according to data augmentation, transfer learning, and interactive segmentation, separately. Compared to existing surveys, this survey classifies the literatures very differently from before and is more convenient for readers to understand the relevant rationale and will guide them to think of appropriate improvements in medical image segmentation based on deep learning approaches.
\end{abstract}


\IEEEpeerreviewmaketitle

\section{Introduction}
Medical image segmentation aims to make anatomical or pathological structures changes in more clear in images; it often plays a key role in computer aided diagnosis and smart medicine due to the great improvement in diagnostic efficiency and accuracy. Popular medical image segmentation tasks include liver and liver-tumor segmentation \cite{li2015automatic} \cite{vivanti2015automatic}, brain and brain-tumor segmentation~\cite{menze2014multimodal}~\cite{cherukuri2017learning}, optic disc segmentation ~\cite{cheng2013superpixel}~\cite{fu2018joint}, cell segmentation~\cite{ronneberger2015u}~\cite{song2017dual}, lung segmentation, pulmonary nodules~\cite{wang2017central}~\cite{onishi2020multiplanar}, cardiac image segmentation~\cite{wu2020cf}~\cite{chen2020deep}, etc. With the development and popularization of medical imaging equipments, X-ray, Computed Tomography (CT), Magnetic Resonance Imaging (MRI) and ultrasound have become four important image assisted means to help clinicians diagnose diseases, to evaluate prognopsis, and to plan operations in medical institutions. In practical applications, although these ways of imaging have advantages as well as disadvantages, they are useful for the medical examination of different parts of human body.

To help clinicians make accurate diagnosis, it is necessary to segment some crucial objects in medical images and extract features from segmented areas. Early approaches to medical image segmentation often depend on edge detection, template matching techniques, statistical shape models, active contours, and machine learning, etc. Zhao et al.~\cite{yu2006medical} proposed a new mathematical morphology edge detection algorithm for lung CT images. Lalonde et al.~\cite{lalonde2001fast} applied Hausdorff-based template matching to disc inspection, and Chen et al.~\cite{chen2009automated} also employed template matching to perform ventricular segmentation in brain CT images. Tsai et al.~\cite{tsai2003shape} proposed a shape based approach using horizontal sets for 2D segmentation of cardiac MRI images and 3D segmentation of prostate MRI images. Li et al.~\cite{li2013likelihood} used the activity profile model to segment liver-tumors from abdominal CT images, while Li et al.~\cite{li2004svm}  proposed a framework for medical body data segmentation by combining level sets and support vector machines (SVMs). Held et al.~\cite{held1997markov} applied Markov random fields (MRF) to brain MRI image segmentation. Although a large number of approaches have been reported and they are successful in certain circumstances, image segmentation is still one of the most challenging topics in the field of computer vision due to the difficulty of feature representation. In particular, it is more difficult to extract discriminating features from medical images than normal RGB images since the former often suffers from problems of blur, noise, low contrast, etc. Due to the rapid development of deep learning techniques~\cite{krizhevsky2017imagenet}, medical image segmentation will no longer require hand-crafted feature and convolutional neural networks (CNN) successfully achieve hierarchical feature representation of images, and thus become the hottest research topic in image processing and computer vision. As CNNs used for feature learning are insensitive to image noise, blur, contrast, etc., they provide excellent segmentation results for medical images.

It is worth mentioning that there are currently two categories of image segmentation tasks, semantic segmentation and instance segmentation. Image semantic segmentation is a pixel-level classification that assigns a corresponding category to each pixel in an image. Compared to semantic segmentation, the instance segmentation not only needs to achieve pixel-level classification, but also needs to distinguish instances on the basis of specific categories. In fact, there are few reports on instance segmentation in medical image segmentation since each organ or tissue is quite different. In this paper, we review the advances of deep learning techniques on medical image segmentation.

\begin{figure}[ht]
	\centerline{\includegraphics[width=\columnwidth]{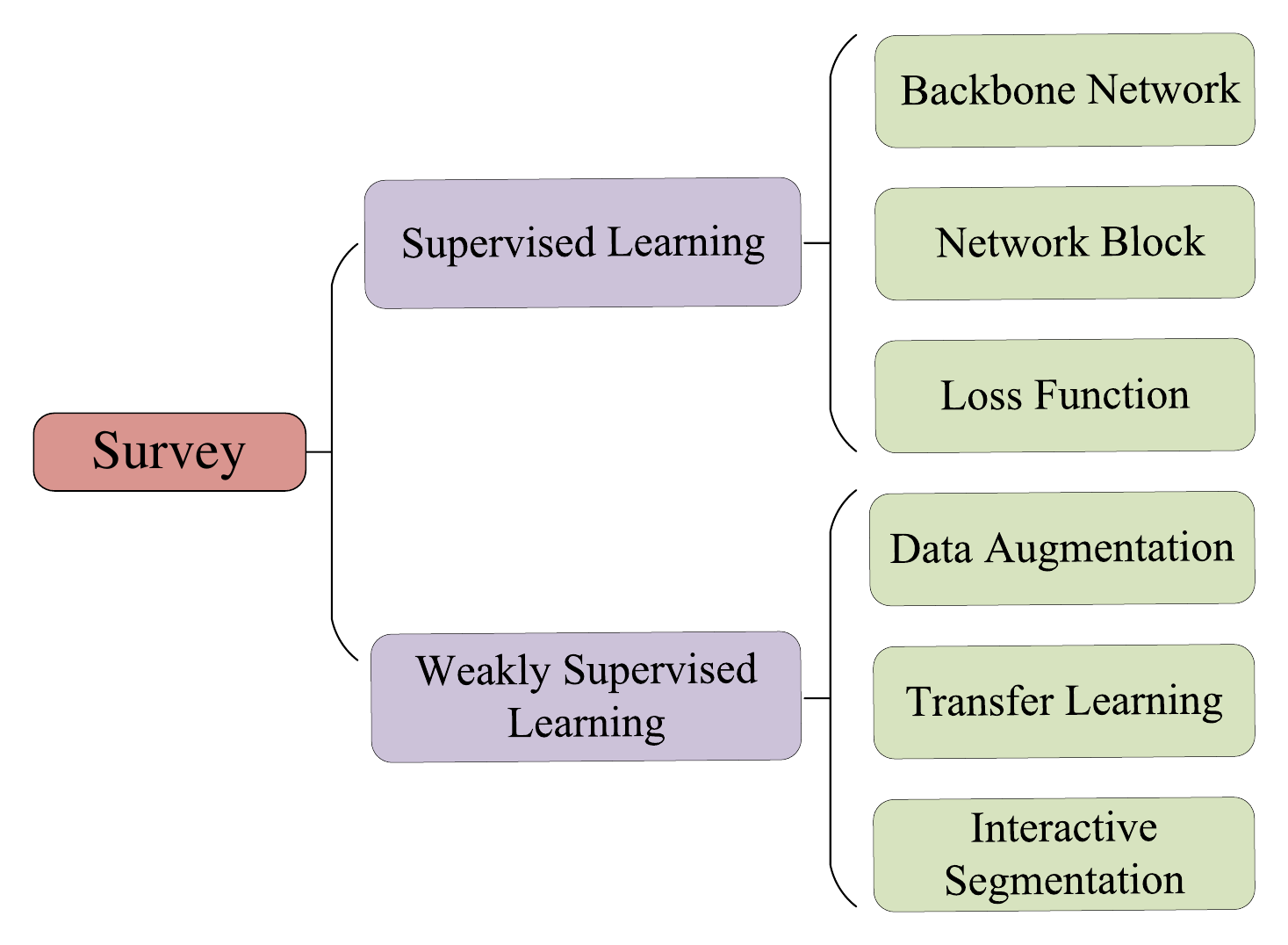}}
	\caption{An overview of deep learning methods on medical image segmentation}
	\label{fig1}
\end{figure}

According to the number of labeled data, machine learning is often categorized into supervised learning, weakly supervised learning, and unsupervised learning. The advantage of supervised learning is that we can train models based on carefully labeled data, but it is difficult to obtain a large number of labeled data for medical images. On the contrary, labeled data are not required for unsupervised learning, but the difficulty of learning is increased. Weakly supervised learning is between the supervised and unsupervised learning since it only requires a small part of data labeled while most of data are unlabeled.

Prior to the widespread application of deep learning, researchers had presented many approaches based on model-driven on medical image segmentation. Masood et al.~\cite{masood2015survey}, made a comprehensive summary of many model-driven techniques in medical image analysis, including image clustering, region growing, and random forest. In~\cite{masood2015survey}, authors summarized different segmentation approaches on medical images according to different mathematical models. Recently, only a few studies based on model-driven techniques were reported, but more and more studies based on data-driven  were reported for medical image segmentation. In this paper, we mainly focus on the evolution and development of deep learning models on medical image segmentation.

In~\cite{shen2017deep}, Shen et al. presented a special review of the application of deep learning in medical image analysis. This review summarizes the progress of machine learning and deep learning in medical image registration, anatomy and cell structure detection, tissue segmentation, computer-aided disease diagnosis and prognopsis. Litjens et al.~\cite{litjens2017survey}reported a survey of deep learning methods, the survey covers the use of deep learning in image classification, object detection, segmentation, registration and other tasks.

More recently, Taghanaki et al.~\cite{taghanaki2020deep}discussed the development of semantic and medical image segmentation; they categorized deep learning-based image segmentation solutions into six groups, i.e., deep architectural, data synthesis-based, loss function-based, sequenced models, weakly supervised, and multi-task methods. To develop a more complete survey on medical image segmentation, Seo et al.~\cite{seo2020machine}reviewed classical machine learning algorithms such as Markov random fields, $k$-means clustering, random forest, and reviewed latest deep learning architectures such as the artificial neural networks (ANNs), the convolutional neural networks (CNNs), the recurrent neural networks (RNNs), etc. Tajbakhsh et al.~\cite{tajbakhsh2020embracing}reviewed solutions of medical image segmentation with imperfect datasets, including two major dataset limitations: scarce annotations and weak annotations. All these surveys play an important role for the development of medical image segmentation techniques. Hesamian et al. ~\cite{hesamian2019deep}reviewed on three aspects of approaches (network structures), training techniques, and challenges. The network structures section describes the main, popular network structures used for image segmentation. The training techniques section discusses the J Digit imaging technique used to train deep neural network models. The challenges section describes the various challenges associated with medical image segmentation using deep learning techniques. Meyer et al.~\cite{meyer2018survey}reviewed the advances in the application or potential application of deep learning to radiotherapy. Akkus et al.~\cite{akkus2017deep}provided an overview of current deep learning-based segmentation approaches for quantitative brain MRI images. Zhou et al.~\cite{zhou2018brief}focused on three typical types of weak supervision: incomplete supervision, inexact supervision and inaccurate supervision. Eelbode et al.~\cite{eelbode2020optimization}focus on evaluating and summarizing the optimization methods used in medical image segmentation tasks based primarily on Dice scores or Jaccard indices.

Through studying the aforementioned surveys, researchers can learn the latest techniques of medical image segmentation, and then make more significant contributions for computer aided diagnoses and smart healthcare. However, these surveys suffer from two problems. One is that most of them chronologically summarize the development of medical image segmentation, and they thus ignore the technical branch of deep learning for medical image segmentation. The other problem is that these surveys only introduce related technical development but not focus on the task characteristics of medical image segmentation such as few-shot learning, imbalance learning, etc., which limits the improvement of medical image segmentation based on task-driven. To address these two problems, we present a novel survey on medical image segmentation using deep learning. In this work, we make the following contributions:

1. We summarize the technical branch of deep learning for medical image segmentation from coarse to fine as shown in Fig. 1. The summation includes two aspects of supervised learning and weakly supervised learning. The latest applications of neural architecture search (NAS), graph convolutional networks (GCN), multi-modality data fusion and medical transformer in medical image analysis are also discussed. Compared to the previous surveys, our survey follows conceptual developments and is believed to be clearer.

2. On supervised learning approaches we analyze literature from three aspects: the selection of backbone networks, the design of network blocks, and the improvement of loss functions. This classification method can help subsequent researchers to understand more deeply motivations and improvement strategies of medical image segmentation networks. For weakly supervised learning, we also review literatures from three aspects for processing few-shot data or class imbalanced data: data augmentation, transfer learning, and interactive segmentation. This organization is expected to be more conducive to researchers in finding innovations for improving the accuracy of medical image segmentation.

3. In addition to reviewing comprehensively the development and application of deep learning in medical image segmentation, we also collect the currently common public medical image segmentation datasets. Finally, we discuss future research trends and directions in this field.

The rest of this paper is organized as follows. In Section \uppercase\expandafter{\romannumeral2}, we review the development and evolution of supervised learning applied to medical images, including the selection of backbone network, the design of network blocks, and the improvement of loss function. In Section \uppercase\expandafter{\romannumeral3}, we introduce the application of unsupervised or weakly supervised methods in the field of medical image segmentation and analyze the commonly unsupervised or weakly supervised strategies for processing few-shot data or class imbalanced data. In Section \uppercase\expandafter{\romannumeral4}, we briefly introduce some of the most advanced methods of medical image segmentation, including NAS, application of GCN, multi-modality data fusion, etc. In Section \uppercase\expandafter{\romannumeral5}, we collect the currently available public medical image segmentation datasets, and summarize limitations of current deep learning methods and future research directions.

\section{Supervised learning}
For medical image segmentation tasks, supervised learning is the most popular method since these tasks usually require high accuracy. In this section, we focus on the review of improvements of neural network architectures. These improvements mainly include network backbones, network blocks and the design of loss functions. Fig. 2 shows an overview on the improvement of network architectures based on supervised learning.
\begin{figure}[htbp]
	\centerline{\includegraphics[width=\columnwidth]{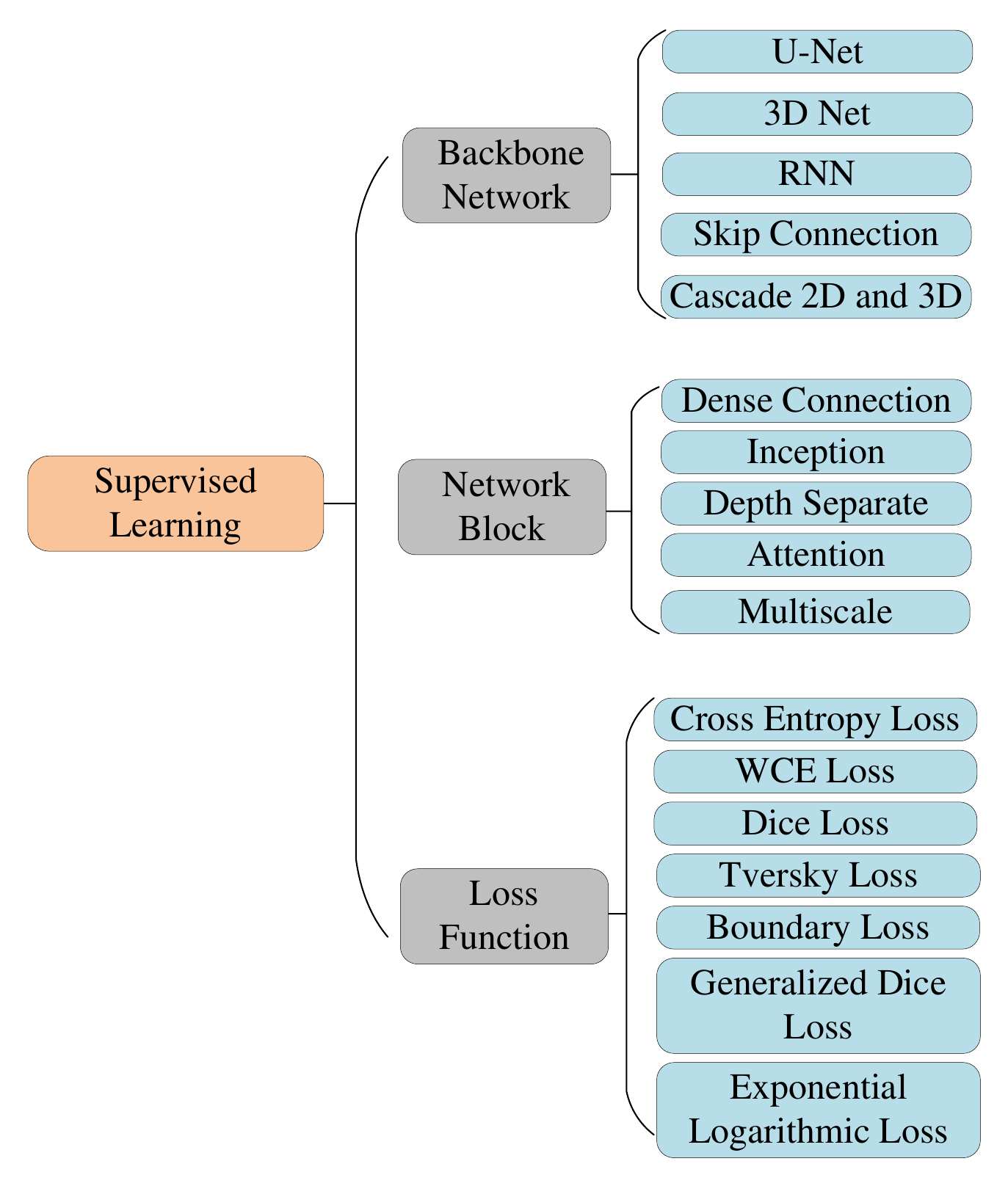}}
	\caption{An overview of network architectures based on supervised learning.}
	\label{fig2}
\end{figure}

\subsection{Backbone Networks}
Image semantic segmentation aims to achieve pixel classification of an image. For this goal, researchers proposed the encoder-decoder structure that is one of the most popular end-to-end architectures, such as fully convolution network (FCN)~\cite{long2015fully}, U-Net~\cite{ronneberger2015u},Deeplab~\cite{chen2017rethinking}, etc. In these structures, an encoder is often used to extract image features while a decoder is often used to restore extracted features to the original image size and output the final segmentation results. Although the end-to-end structure is pragmatic for medical image segmentation, it reduces the interpretability of models. The first high-impact encoder-decoder structure, the U-Net proposed by Ronneberger et al.~\cite{ronneberger2015u}has been widely used for medical image segmentation. Fig. 3 shows the U-Net architecture.

\emph{U-Net:} The U-Net solves problems of general CNN networks used for medical image segmentation, since it adopts a perfect symmetric structure and skip connection. Different from common image segmentation, medical images usually contain noise and show blurred boundaries. Therefore, it is very difficult to detect or recognize objects in medical images only depending on image low-level features. Meanwhile, it is also impossible to obtain accurate boundaries depending only on image semantic features due to the lack of image detail information. Whereas, the U-Net effectively fuses low-level and high-level image features by combining low-resolution and high-resolution feature maps through skip connections, which is a perfect solution for medical image segmentation tasks. Currently, the U-Net has become the benchmark for most medical image segmentation tasks and has inspired a lot of meaningful improvements.

\begin{figure}[htbp]
	\centerline{\includegraphics[width=\columnwidth]{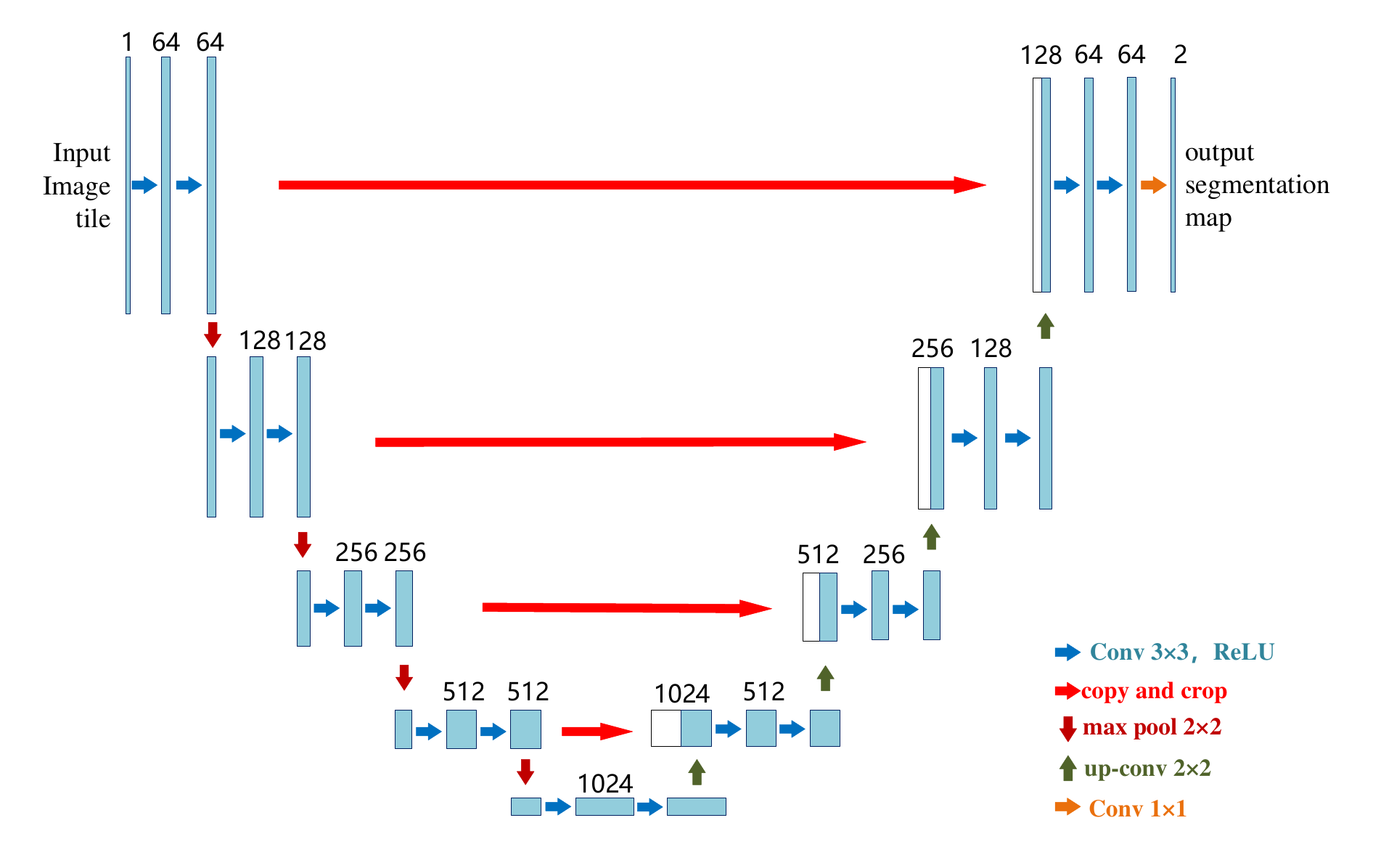}}
	\caption{The U-Net architecture~\cite{ronneberger2015u}.}
	\label{fig3}
\end{figure}

\emph{3D Net:} In practice, as most of medical data such as CT and MRI images exist in the form of 3D volume data, the use of 3D convolution kernels can better mine the high-dimensional spatial correlation of data. Motivated by this idea, Çiçek et al.~\cite{cciccek20163d}extended U-Net architecture to the application of 3D data, and proposed 3D U-Net that deals with 3D medical data directly. Due to the limitation of computational resources, the 3D U-Net only includes three down-sampling, which cannot effectively extract deep-layer image features leading to limited segmentation accuracy for medical images. In addition, Milletari et al.~\cite{milletari2016v}proposed a similar architecture, V-Net, as shown in Fig. 4. It is well known that residual connections can avoid vanishing gradient and accelerate network convergence, and it is thus easy to design deeper network structures that can provide better feature representation. Compared to 3D U-Net, V-Net employs residual connections to design a deeper network (4 down-samplings), and thus achieves higher performance. Similarly, by applying residual connections to 3D networks, Yu et al.~\cite{chen2016voxresnet}presented Voxresnet, Lee et al.~\cite{lee2017superhuman}proposed 3DRUNet, and Xiao et al.~\cite{xiao2018weighted}proposed Res-UNet. However, these 3D Networks encounter same problems of high computational cost and GPU memory usage due to a very large number of parameters.

\begin{figure}[htbp]
	\centerline{\includegraphics[width=\columnwidth]{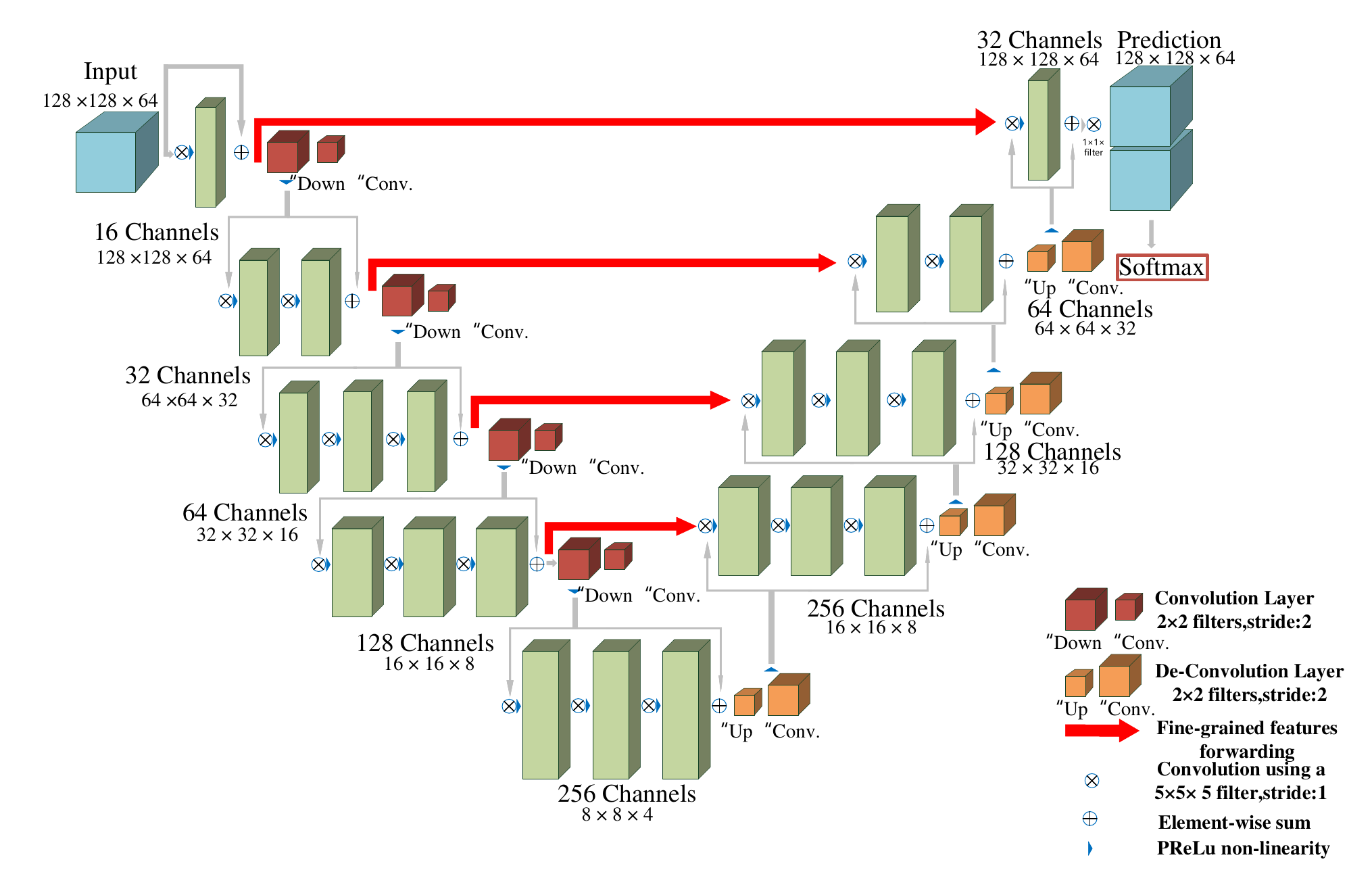}}
	\caption{The V -Net architecture~\cite{milletari2016v}.}
	\label{fig4}
\end{figure}

\emph{Recurrent Neural Network (RNN):} RNN is initially designed to deal with sequence problems. The long Short-Term Memory (LSTM) network~\cite{hochreiter1997long}is one of the most popular RNNs. It can retain the gradient flow for a long time by introducing a self-loop. For medical image segmentation, RNN has been used to model the time dependence of image sequences. Alom et al.~\cite{alom2018recurrent}proposed a medical image segmentation method that combines ResUNet with RNN. The method achieves feature accumulation of recursive residual convolutional layers, which improves feature representation for image segmentation tasks. Fig. 5 shows the recurrent residual convolutional unit. Gao et al.~\cite{gao2018fully}joined LSTM and CNN to model the temporal relationship between different brain MRI slices to improve segmentation accuracy. Bai et al.~\cite{bai2018recurrent}combined FCN with RNN to mine the spatiotemporal information for aortic sequence segmentation. Clearly, RNN can capture local and global spatial features of images by considering the context information relationship. However, in medical image segmentation, the capture of complete and valid temporal information requires good medical image quality (e.g. smaller slice thickness and pixel spacing). Therefore, the design of RNN is uncommon for improving the performance of medical image segmentation.

\begin{figure}[ht]
	\centerline{\includegraphics[width=\columnwidth]{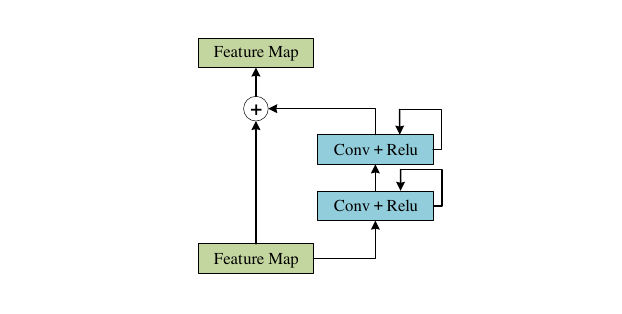}}
	\caption{The recurrent residual convolutional unit{alom2018recurrent}.}
	\label{fig5}
\end{figure}

\emph{Skip Connection:} Although the skip connection can fuse low-resolution and high-resolution information and thus improve feature representation, it suffers from the problem of the large semantic gap between low- and high-resolution features, leading to blurred feature maps. To improve skip connection, Ibtehaz et al.~\cite{ibtehaz2020multiresunet}proposed MultiResUNet including the Residual Path (ResPath), which makes the encoder features perform some additional convolution operations before fusing with the corresponding features in the decoder. Seo et al.~\cite{seo2019modified} proposed mUNet and Chen et al.~\cite{chen2019feature}proposed FED-Net. Both mU-Net and FED-Net add convolution operations to the skip connection to improve the performance of medical image segmentation.

\emph{Cascade of 2D and 3D:} For image segmentation tasks, the cascade model often trains two or more models to improve segmentation accuracy. This method is especially popular in medical image segmentation. The cascade model can be broadly divided into three types of frameworks: coarse-fine segmentation, detection  segmentation, and mixed segmentation. The first class is a coarse-fine segmentation framework that uses a cascade of two 2D networks for segmentation, where the first network performs coarse segmentation and then uses another network model to achieve fine segmentation based on the previous coarse segmentation results. Christ et al.~\cite{christ2016automatic}proposed a cascaded network for liver and liver-tumor segmentation. This network firstly uses a FCN to segment livers, and then uses previous liver segmentation results as the input of the second FCN for liver-tumor segmentation. Yuan et al.~\cite{tang2018dsl}first trained a simple convolutional-deconvolutional neural networks (CDNN) model (19-layer FCN) to provide rapid but coarse liver segmentation over the entire images of a CT volume, and then applied another CDNN (29-layer FCN) to the liver region for fine-grained liver segmentation. Finally, the liver segmentation region enhanced by histogram equalization is considered as an additional input to the third CDNN (29-layer CNN) for liver-tumor segmentation. Besides, other networks using the coarse-fine segmentation framework can be found in~\cite{kaluva20182d}~\cite{feng2019automatic}~\cite{albishri2019cu}. At the same time, the detection segmentation framework is also popular. First, a network model such as R-CNN~\cite{he2017mask}or You-On-Look-Once (YOLO)~\cite{bochkovskiy2020yolov4}is used for target location identification, and then another network is used for further detailed segmentation based on previously coarse segmentation results. Al-Antari et al.~\cite{al2018fully}proposed a similar approach for breast mass detection, segmentation and classification from mammograms. In this work, the first step is to use the regional deep learning method YOLO for target detection, the second step is to input the detected targets into a newly designed full-resolution convolutional network (FrCN) for segmentation, and finally, a deep convolutional neural network is used to identify the masses and classify them as benign or malignant. Similarly, Tang et al.~\cite{tang2018dsl}used faster R-CNN~\cite{ren2016faster}and Deeplab~\cite{chen2017deeplab}cascades for localization segmentation of the liver. In addition, both Salehi et al.~\cite{salehi2017auto}and Yan et al.~\cite{yan2019cascaded}proposed a kind of cascade networks for whole-brain MRI and high-resolution mammogram segmentation. This kind of cascade network can effectively extract richer multi-scale context information by using a posteriori probabilities generated by the first network than normal cascade networks.

However, most of medical images are 3D volume data, but a 2D convolutional neural network cannot learn temporal information in the third dimension, and a 3D convolutional neural network often requires high computation cost and severes GPU memory consumption. Therefore some pseudo-3D segmentation methods have been proposed. Oda et al.~\cite{oda2019abdominal}proposed a three-plane method of cascading three networks to segment the abdominal artery region effectively from the medical CT volume. Vu et al.~\cite{vu2019evaluation}applied the overlay of adjacent slices as input to the central slice prediction, and then fed the obtained 2D feature map into a standard 2D network for model training. Although these pseudo-3D approaches can segment object from 3D volume data, they only obtain limited accuracy improvement due to the utilization of local temporal information. Compared to pseudo-3D networks, hybrid cascading 2D and 3D networks are more popular. Li et al.~\cite{li2018h}proposed a hybrid densely connected U-Net (H-DenseUNet) for liver and liver-tumor segmentation. This method first employs a simple Resnet to obtain a rough liver segmentation result, utilizing the 2D DenseUNet to extract 2D image features effectively, then uses the 3D DenseUNet to extract 3D image features, and finally designs a hybrid feature fusion layer to jointly optimize 2D and 3D features. Although the H-DenseUNet reduces the complexity of models compared to an entire 3D network, the model is complex and it still suffers from a large number of parameters from 3D convolutions. For the problem, Zhang et al.~\cite{zhang2019light}proposed a lightweight hybrid convolutional network (LW-HCN) with a similar structure to the H-DenseUNet, but the former requires fewer parameters and computational cost than the latter due to the design of the depthwise and spatiotemporal separate (DSTS) block and the use of 3D depth separable convolution. Similarly, Dey et al.~\cite{dey2020hybrid}also designed a cascade of 2D and 3D network for liver and liver-tumor segmentation.

Obviously, among the three types of cascade networks mentioned above, the hybrid 2D and 3D cascade network can effectively improve segmentation accuracy and reduce the learning burdens.

In contrast to the above cascade networks, Valanarasu et al.~\cite{valanarasu2020kiu}proposed a complete cascade network namely KiU-Net to perform brain dissection segmentation. The performance of vanilla U-Net is greatly degraded when detecting smaller anatomical structures with fuzzy noise boundaries. To overcome this problem, authors designed a novel over-complete architecture Ki-Net, in which the spatial size of the intermediate layer is larger than that of the input data, and this is achieved by using an up-sampling layer after each conversion layer in the encoder. Thus the proposed Ki-Net possesses stronger edge capture capability compared to U-Net and finally it is cascaded with the vanilla U-Net to improve the overall segmentation accuracy. Since the KiU-Net can exploit both the low-level fine edges feature maps using Ki-Net and the high-level shape feature maps using U-Net, it not only improves segmentation accuracy but also achieves fast convergence for small anatomical landmarks and blurred noisy boundaries.

\emph{Others:}A generating adversarial networks (GAN)~\cite{goodfellow2014generative}has been widely used in many areas of computer vision. In its infancy, the GAN was often used for data augmentation by generating new samples, which would be reviewed in Section \uppercase\expandafter{\romannumeral3}, but later researchers discovered that the idea of generative confrontation could be used in almost any field, and was therefore also used for image segmentation. Since medical images usually show low contrast, blurred boundaries between different tissues or between tissues and lesions, and sparse medical image data with labels, U-Net-based segmentation methods using pixel loss to learn local and global relationships between pixels are not sufficient for medical image segmentation, and the use of generative adversarial networks is becoming a popular idea for improving image segmentation. Luc et al.~\cite{luc2016semantic}firstly applied the generative adversarial network to image segmentation, where the generative network is used for segmentation models and the adversarial network is trained as a classifier. Singh et al.~\cite{singh2020breast}proposed a conditional generation adversarial network (cGAN) to segment breast tumors within the target area (ROI) in mammograms. The generative network learns to identify tumor regions and generates segmentation results, and the adversarial network learns to distinguish between ground truth and segmentation results from the generative network, thereby enforcing the generative network to obtain labels as realistic as possible. The cGAN works fine when the number of training samples is limited. Conze et al.~\cite{conze2020abdominal}utilized cascaded pretrained convolutional encoder-decoders as generators of cGAN for abdominal multi-organ segmentation, and considered the adversarial network as a discriminator to enforces the model to create realistic organ delineations.

In addition, the incorporation of the prior knowledge about organ shape and position may be crucial for improving medical image segmentation effect, where images are corrupted and thus contain artefacts due to limitations of imaging techniques. However, there are few works about how to incorporate prior knowledge into CNN models. As one of the earliest studies in this field, Oktay et al.~\cite{oktay2017anatomically}proposed a novel and general method to combine a priori knowledge of shape and label structure into the anatomically constrained neural networks (ACNN) for medical image analysis tasks. In this way, the neural network training process can be constrained and guided to make more anatomical and meaningful predictions, especially in cases where input image data is not sufficiently informative or consistent enough (e.g., missing object boundaries). Similarly, Boutillon et al.~\cite{boutillon2020combining}incorporated anatomical priors into a conditional adversarial framework for scapula bone segmentation, combining shape priors with conditional neural networks to encourage models to follow global anatomical properties in terms of shape and position information, and to make segmentation results as accurate as possible. The above study shows that improved models can provide higher segmentation accuracy and they are more robust since priori knowledge constraints are employed in the training process of neural networks.

After proposing U-Net in~\cite{ronneberger2015u}, the encoder-decoder structure became the most popular structure in medical image segmentation. The design of the network backbone focuses on more efficient feature extraction in the encoder and feature recovery and fusion in the decoder to improve segmentation accuracy.

\subsection{Network Function Block}
\subsubsection{Dense Connection}
Dense connection is often used to construct a kind of special convolution neural networks. For dense connection networks, the input of each layer comes from the output of all previous layers in the process of forward transmission. Inspired by the dense connection, Guan et al. ~\cite{guan2019fully} proposed an improved U-Net by replacing each sub-block of U-Net with a form of dense connections as shown in Fig. 6. Although the dense connection is helpful for obtaining richer image features, it often reduces the robustness of feature representation to a certain extent and increases the number of parameters.

Zhou et al. ~\cite{zhou2019unet++} connected all U-Net layers (from one to four) together as shown in Fig. 7. The advantage of this structure is that it allows the network to learn automatically importance of features at different layers. Besides, the skip connection is redesigned so that features with different semantic scales can be aggregated in the decoder, resulting in a highly flexible feature fusion scheme. The disadvantage is that the number of parameters is increased due to the employment of dense connection. Therefore, a pruning method is integrated into model optimization to reduce the number of parameters. Meanwhile, the deep supervision ~\cite{lee2015deeply} is also employed to balance the decline of segmentation accuracy caused by the pruning. Although the dense connection is helpful for obtaining richer image features, it often reduces the robustness of feature representation to a certain extent and increases the number of parameters.

\begin{figure}[htbp]
	\centerline{\includegraphics[width=\columnwidth]{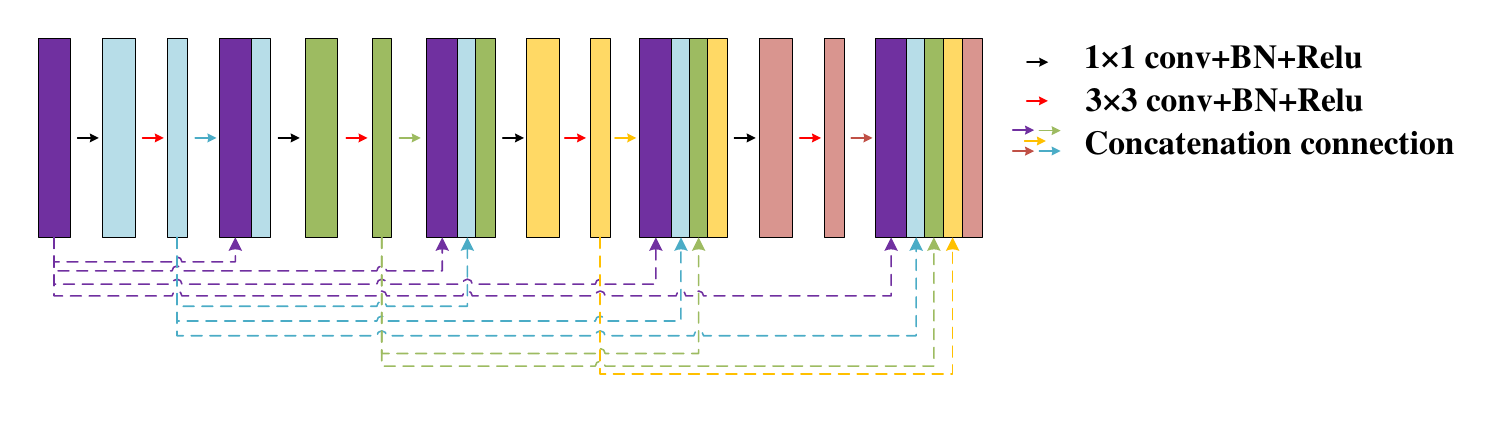}}
	\caption{Dense connection architecture ~\cite{guan2019fully}.}
	\label{fig6}
\end{figure}

\begin{figure}[ht]
	\centerline{\includegraphics[width=\columnwidth]{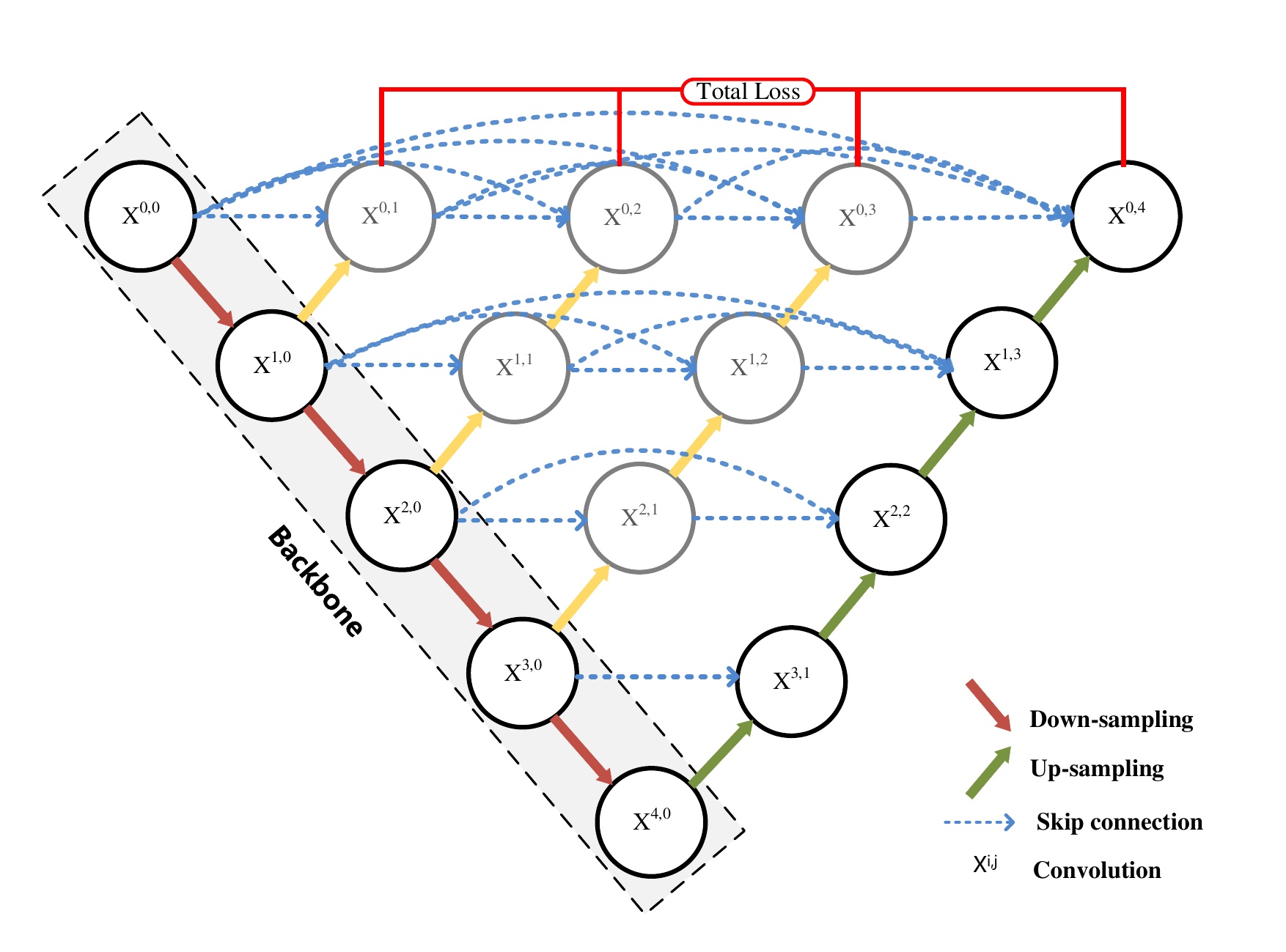}}
	\caption{The U-Net++ architecture ~\cite{zhou2019unet++}.}
	\label{fig7}
\end{figure}
\subsubsection{Inception}
For CNNs, deep networks often give better performances than shallow ones, but they encounter some new problems such as vanishing gradient, the difficulty of network convergence, the requirement of large memory usage, etc. The inception structure overcomes these problems. It gives better performance by merging convolution kernels in parallel without increasing the depth of networks. This structure is able to extract richer image features using multi-scale convolution kernels, and to perform feature fusion to obtain better feature representation. Inspired by GoogleNet ~\cite{szegedy2015going}~\cite{szegedy2016rethinking}, Gu et al. ~\cite{gu2019net} proposed CE-Net by introducing the inception structure into medical image segmentation. The CE-Net adds atrous convolution to each parallel structure to extract features on a wide reception field, and adds $1\times1$ convolution of feature maps, Fig. 8 shows the architecture of the inception. However, the inception structure is complex leading to the difficulty of model modification.

\begin{figure}[ht]
	\centerline{\includegraphics[width=\columnwidth]{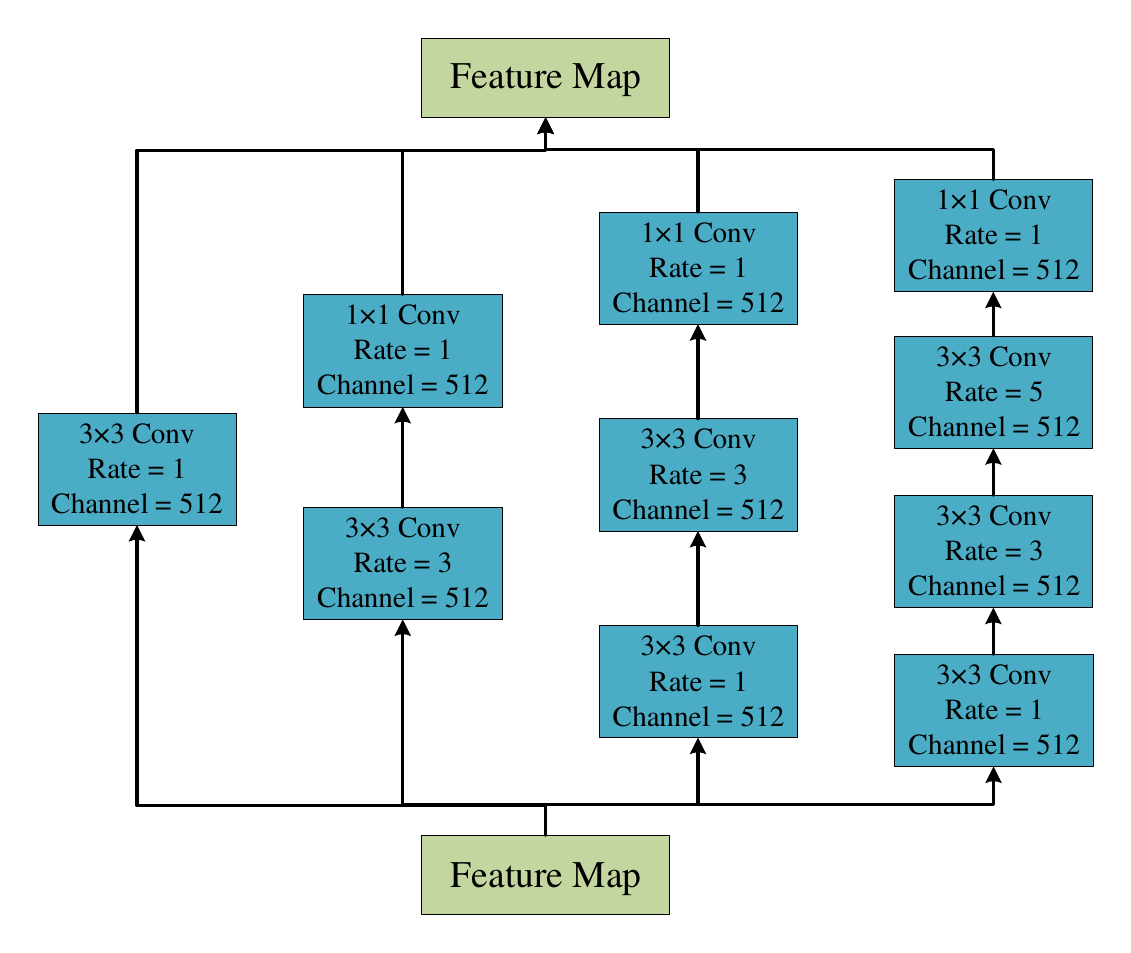}}
	\caption{The inception architecture ~\cite{gu2019net}. It contains four cascade branches with the gradual increment of the number of atrous convolution, from 1 to 1, 3, and 5, then the receptive field of each branch will be 3, 7, 9, and 19. Therefore, the network can extract features from different scales.}
	\label{fig8}
\end{figure}

\subsubsection{Depth Separability}
To improve the generalization capability of network models and to reduce the requirement of memory usage, many researchers focus on the study of lightweight networks for complex medical 3D volume data. Howard et. al. ~\cite{howard2017mobilenets} proposed MobileNet to decompose vanilla convolution into depthwise separable convolution and pointwise convolution. The number of vanilla convolution operation is usually  ${D_K\times D_K\times M\times N}$, where ${M}$ is the dimension of the input feature maps, ${N}$ is the dimension of the output feature maps, ${D_K}$ is the size of the convolution kernels. However, the number of the channel convolution operation is ${D_K\times D_K\times 1\times M}$ and the point convolution is ${1\times 1\times M\times N}$. Compared to vanilla convolution, the computational cost of depthwise separable convolution is (1/$N$ + 1/${D_K^2}$) times than that of the vanilla convolution. Based on this, Sandler et al. ~\cite{sandler2018mobilenetv2} proposed MobileNet-V2 that contains a novel layer module, the inverted residual with linear bottleneck. In this module, the input is a low-dimensional compressed representation which is first expanded to high dimension and then filtered with a lightweight depthwise convolution. Features are subsequently projected back to a low-dimensional representation with a linear convolution. It allows to significantly reduce the memory footprint needed during inference. By extending the depth separable convolution to the design of 3D networks, Lei et al. ~\cite{lei2020lightweight} proposed a lightweight V-Net (LV-Net) with fewer operations than V-Net for liver segmentation.Besides, Zhang et al. ~\cite{zhang2019light} and Huang et al. ~\cite{huang20193d} also proposed the application of depthwise separable convolutions to the segmentation of 3D medical volume data. Other related works for lightweight deep networks can be found in ~\cite{paschali20193dq}~\cite{xu2018quantization}. Depthwise separable convolution is an effective way to reduce the number of model parameters, but it may result in loss of accuracy in medical image segmentation, and thus other approaches (e.g. deep supervision)~\cite{lei2020lightweight} need to be employed to improve segmentation accuracy.

\subsubsection{Attention Mechanism}
For neural networks, an attention block can selectively change input or assigns different weights to input variables according to different importance. In recent years, most of researches combining deep learning and visual attention mechanism have focused on using masks to form attention mechanisms. The principle of masks is to design a new layer that can identify key features from an image, through training and learning, and then let networks only focus on interesting areas of images.

\emph{Local Spatial Attention:} The spatial attention block aims to calculate the feature importance of each pixel in space-domain and extract the key information of an image. Jaderberg et al. ~\cite{jaderberg2015spatial} early proposed a spatial transformer network (ST-Net) for image classification by using spatial attention that transforms the spatial information of an original image into another space and retains the key information. Normal pooling is equivalent to the information merge that easily causes the loss of key information. For this problem, a block called spatial transformer is designed to extract key information of images by performing a spatial transformation. Inspired by this, Oktay et al. ~\cite{oktay2018attention} proposed attention U-Net. The improved U-Net uses an attention block to change the output of the encoder before fusing features from the encoder and the corresponding decoder. The attention block outputs a gating signal to control feature importance of pixels at different spatial positions. Fig. 9 shows the architecture. This block combines the Relu and sigmoid functions via $1\times1$ convolution to generate a weight map that is corrected by multiplying features from the encoder.

\begin{figure}[htbp]
	\centerline{\includegraphics[width=\columnwidth]{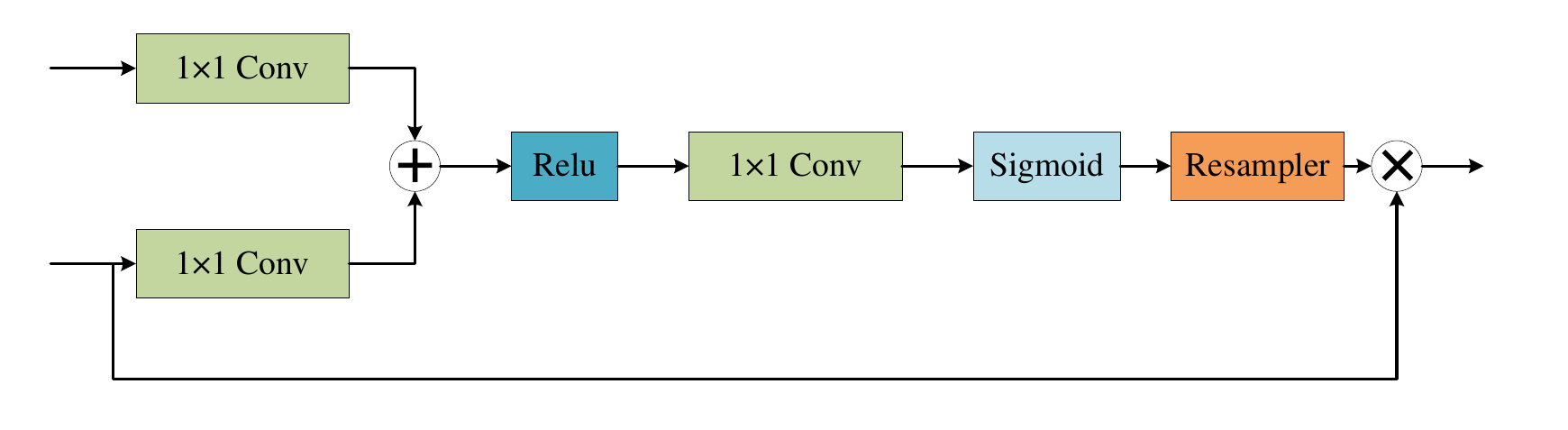}}
	\caption{The attention block in the attention U-Net ~\cite{oktay2018attention}.}
	\label{fig9}
\end{figure}

\emph{Channel Attention:} The channel attention block can achieve feature recalibration, which utilizes learned global information to emphasize selectively useful features and suppress useless features. Hu et al. ~\cite{hu2018squeeze} proposed SE-Net that introduced the channel attention to the field of image analysis and won the ImageNet Challenge in 2017. This method implements attention weighting on channels using three steps; Fig. 10 shows this architecture. The first is the squeezing operation, the global average pooling is performed on input features to obtain the $1\times1\times Channel$ feature map. The second is the excitation operation, where channel features are interacted to reduce the number of channels, and then the reduced channel features are reconstructed back to the number of channels. Finally the sigmoid function is employed to generate a feature weight map of $[0, 1]$ that multiplies the scale back to the original input feature. Chen et al. ~\cite{chen2019feature} proposed FED-Net that uses the SE block to achieve the feature channel attention.
\begin{figure}[htbp]
	\centerline{\includegraphics[width=\columnwidth]{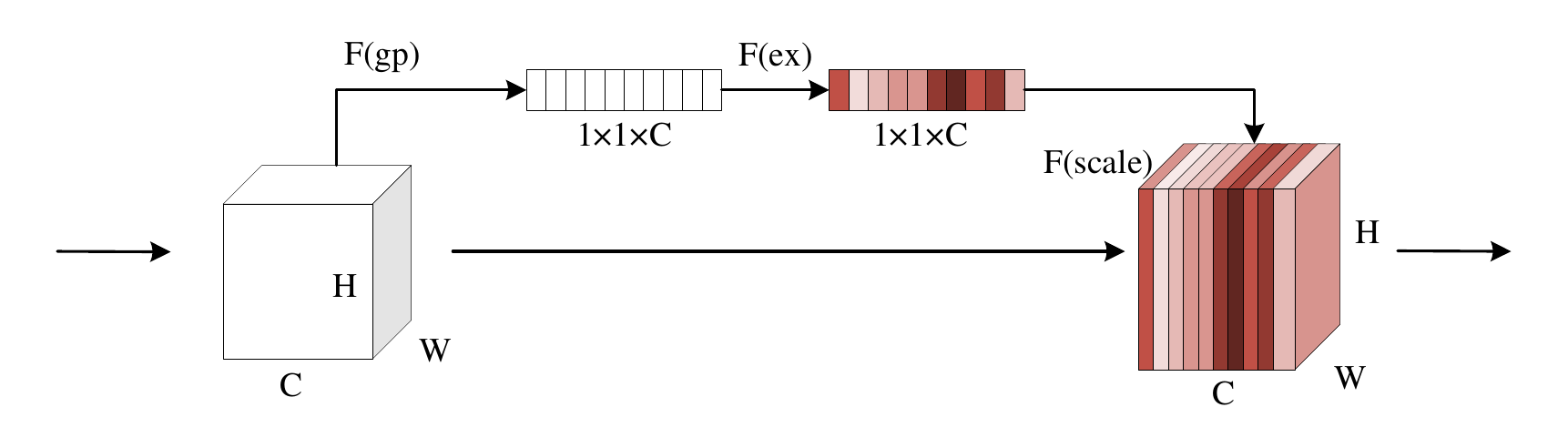}}
	\caption{The channel attention in the SE-Net ~\cite{hu2018squeeze}.}
	\label{fig10}
\end{figure}

\emph{Mixture Attention:} Spatial and channel attention mechanisms are two popular strategies for improving feature representation. However, the spatial attention ignores the difference of different channel information and treats each channel equally. On the contrary, the channel attention pools global information directly while ignoring local information in each channel, which is a relatively rough operation. Therefore, combining advantages of two attention mechanisms, researchers have designed many models based on a mixed domain attention block. Kaul et al. ~\cite{kaul2019focusnet} proposed the focusNet using a mixture of spatial attention and channel attention for medical image segmentation, where the SE-Block is used for channel attention and a branch of spatial attention is designed. Besides, other related works can be found in ~\cite{paschali20193dq}~\cite{xu2018quantization}.

To improve the feature discriminant representation of networks, Wang et al. ~\cite{wang2019sclerasegnet} embedded an attention block inside the central bottleneck between the contraction path and the expansion path of the U-Net, and proposed the ScleraSegNet. Furthermore, they compared the performance of channel attention, spatial attention, and different combinations of two attentions for medical image segmentations. They concluded that the channel-centric attention was the most effective in improving image segmentation performance. Based on this conclusion, they finally won the championship of the sclera segmentation benchmarking competition (SSBC2019).

Although those attention mechanisms mentioned above improve the final segmentation performance, they only perform an operation of local convolution. The operation focuses on the area of neighboring convolution kernels but misses the global information. In addition, the operation of down-sampling leads to the loss of spatial information, which is especially unfavorable for biomedical image segmentation. A basic solution is to extract long-distance information by stacking multiple layers, but this is low efficiency due to a large number of parameters and high computational cost. In the decoder, the up-sampling, the deconvolution, and the interpolation are also performd in the way of local convolution.

\emph{Non-local Attention:} Recently, Wang et al. ~\cite{wang2020non} proposed a Non-local U-Net to overcome the drawback of local convolution for medical image segmentation. The Non-local U-Net employs the self-attention mechanism and the global aggregation block to extract full image information during the parts of both up-sampling and down-sampling, which can improve the final segmentation accuracy. Fig. 11 shows the global aggregation block. The Non-local block is a general-purpose block that can be easily embedded in different convolutional neural networks to improve their performance.

It can be seen that the attention mechanism is effective for improving image segmentation accuracy. In fact, spatial attention looks for interesting target regions while channel attention looks for interesting features. The mixed attention mechanism can take advantages of both spaces and channels. However, compared with the non-local attention, the conventional attention mechanism lacks the ability of exploiting the associations between different targets and features, so CNNs based on non-local attention usually exhibit better performance than normal CNNs for image segmentation tasks.

\begin{figure*}[ht]
	\centerline{\includegraphics[width=\textwidth]{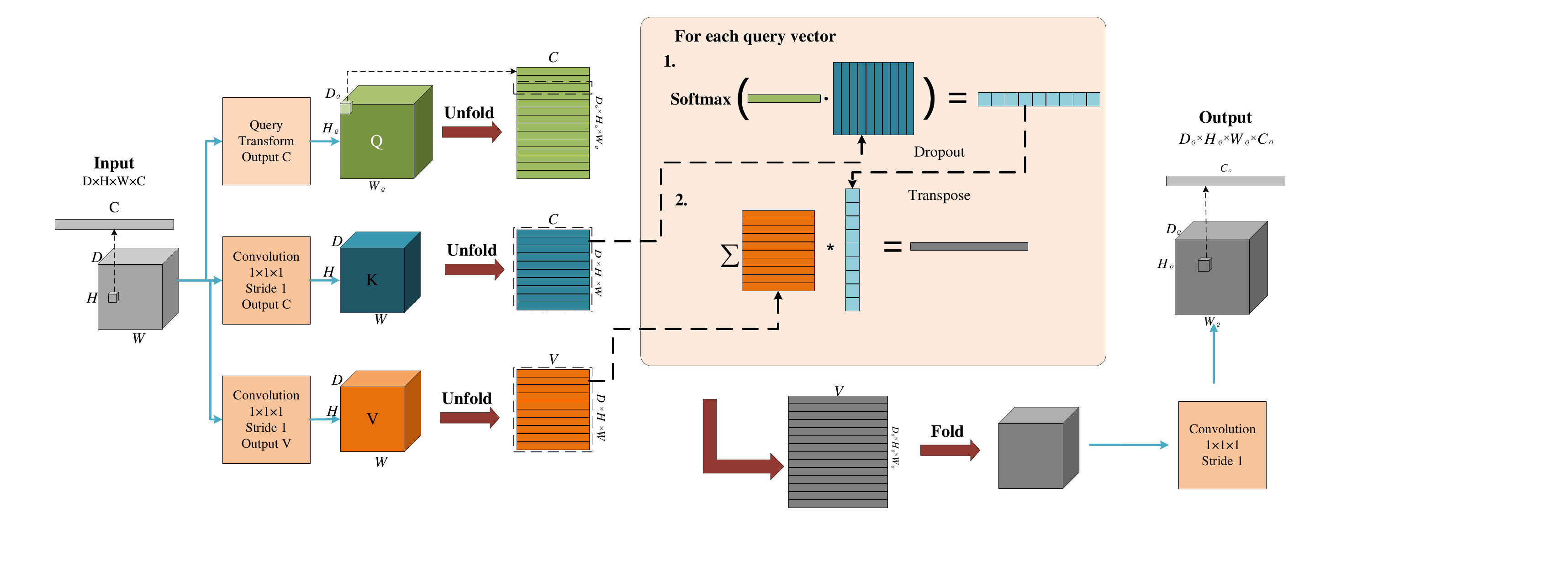}}
	\caption{The global aggregation block in the Non-Local U-Net ~\cite{wang2020non}.}
	\label{fig11}
\end{figure*}

\subsubsection{Multi-scale Information Fusion}
One of the challenges in medical image segmentation is a large range of scales among objects. For example, a tumor in the middle or late stage could be much larger than that in the early stage. The size of perceptive field roughly determines how much context information we can use. The general convolution or pooling only employs a single kernel, for instance, a $3\times3$ kernel for convolution and a $2\times2$ kernel for pooling.

\emph{Pyramid Pooling:} The parallel operation of multi-scale pooling can effectively improve context information of networks, and thus extract richer semantic information. He et al. ~\cite{he2015spatial} first proposed spatial pyramid pooling (SPP) to achieve multi-scale feature extraction. The SPP divides an image from the fine space to the coarse space, then gathers local features and extracts multi-scale features. Inspired by the SPP, a multi-scale information extraction block is designed and named residual multi-kernel pooling (RMP) ~\cite{gu2019net} that uses four pooling kernels with different sizes to encode global context information. However, the up-sampling operation in RMP cannot restore the loss of detail information due to pooling that usually enlarges the receptive field but reduces the image resolution.

\emph{Atrous Spatial Pyramid Pooling:} In order to reduce the loss of detail information caused by pooling operation, researchers proposed atrous convolution instead of the polling operation. Compared with the vanilla convolution, the atrous convolution can effectively enlarge the receptive field without increasing the number of parameters. Combining advantages of the atrous convolution and the SPP block, Chen et al. ~\cite{chen2017deeplab} proposed the atrous spatial pyramid pooling module (ASPP) to improve image segmentation results. The ASPP shows strong recognition capability on same objects with different scales. Similarly, Similarly, Lopez et al ~\cite{lopez2017dilated} and Lei et al ~\cite{9354863} applied superposition of multi-scale atrous convolutions to brain tumor segmentation and liver tumor segmentation, respectively, which achieves a clear accuracy improvement.

However, the ASPP suffers from two serious problems for image segmentation. The first problem is the loss of local information as shown in Fig. 12, where we assume that the convolutional kernel is $3\times3$ and the dilation rate is 2 for three iterations. The second problem is that the information could be irrelevant across large distances. How to simultaneously handle the relationship between objects with different scales is important for designing a fine atrous convolutional network. In response to the above problems, Wang et al. ~\cite{wang2018understanding} designed an hybrid expansion convolution (HDC) networks. This structure uses a sawtooth wave-like heuristic to allocate the dilation rate, so that information from a wider pixel range can be accessed and thus the gridding effect is suppressed. In ~\cite{wang2018understanding}, authors gave several atrous convolution sequences using variable dilation rate, e.g., [1,2,3], [3,4,5], [1,2,5], [5,9,17], and [1,2,5,9].

\begin{figure}[htbp]
	\centerline{\includegraphics[width=\columnwidth]{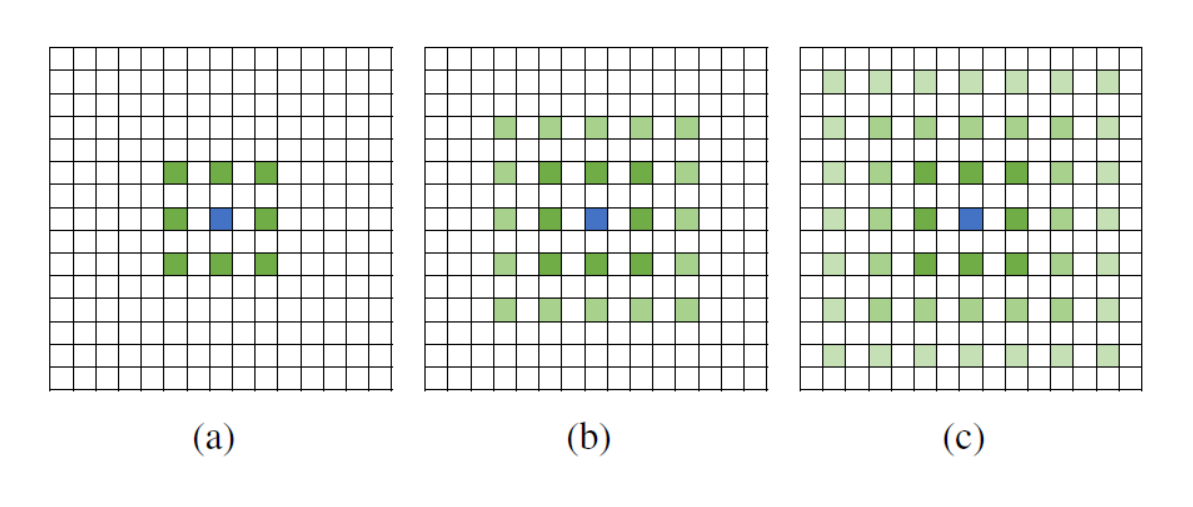}}
	\caption{The gridding effect (the way of treating images as a chessboard causes the loss of information continuity).}
	\label{fig12}
\end{figure}
\emph{Non-local and ASPP:} The atrous convolution can efficiently enlarge the receptive field to collect richer semantic information, but it causes the loss of detail information due to the gridding effect. Therefore, it is necessary to add constraints or establish pixel associations for improving the atrous convolution performance. Recently, Yang et al. ~\cite{yang2019parsing} proposed a combination block of ASPP and Non-local for the segmentation of human body parts, as shown in Fig. 13. ASPP uses multiple parallel atrous convolutions with different scales to capture richer information, and the Non-local operation captures a wide range of dependencies. This combination possesses advantages of both ASPP and Non-local, and it has a good application prospect for medical image segmentation.

\begin{figure}[htbp]
	\centerline{\includegraphics[width=\columnwidth]{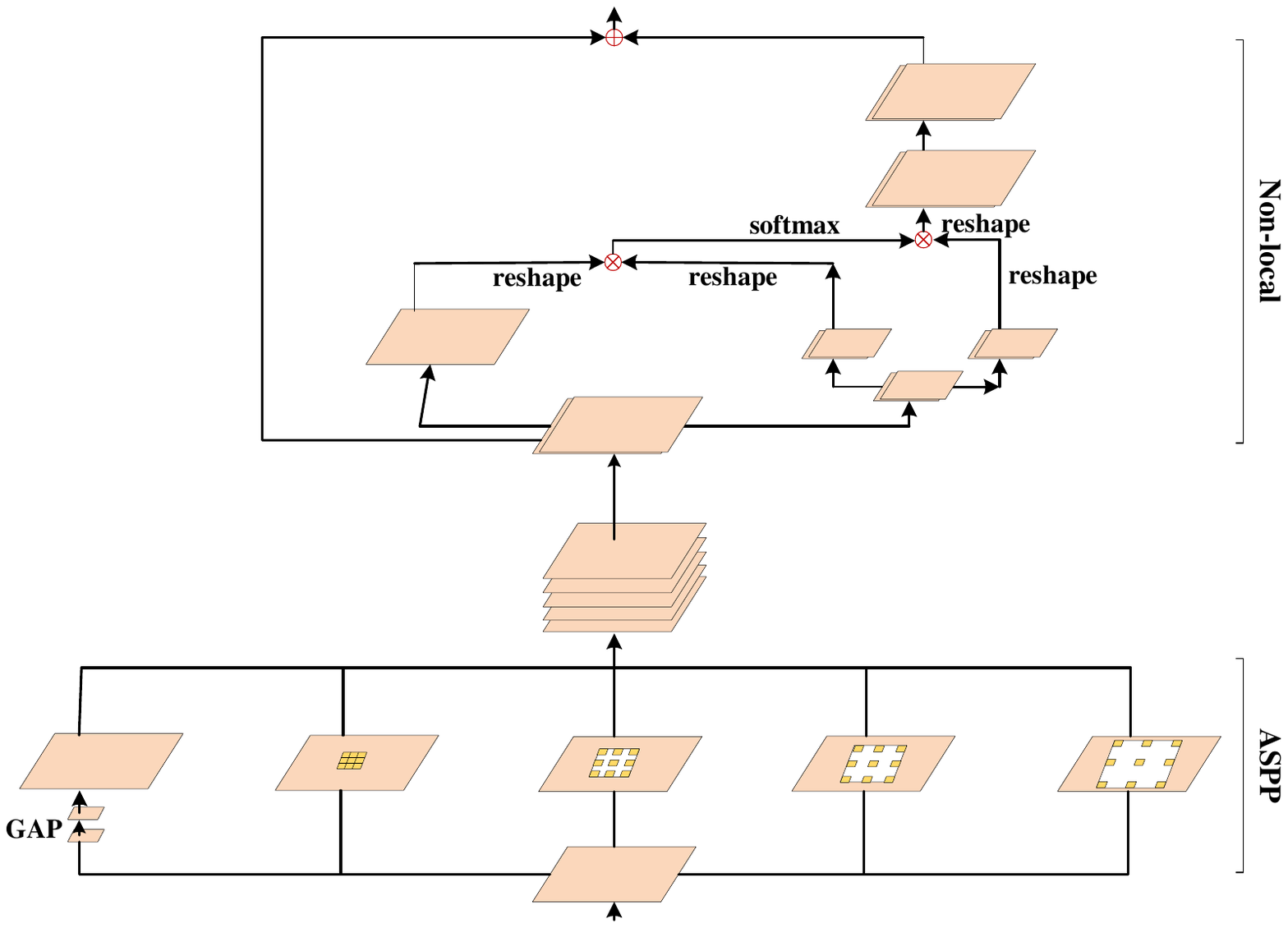}}
	\caption{The combination of ASPP and Non-local architecture ~\cite{yang2019parsing}.}
	\label{fig13}
\end{figure}

The network function module is designed to perform more efficient feature fusion. When feature is usually extracted by the encoder, the feature is usually fused by the network function module to enhance the feature representation. Feature fusion is usually performed by fusing different scale information or performing a more efficient way of feature transfer. Then the feature is passed through the decoder to obtain a better segmentation result.

\subsection{Loss Function}
In addition to improved segmentation speed and accuracy by designing network backbone and the function block, designing new loss functions also resulted in improvements in subsequent inference-time segmentation accuracy. Therefore, a great deal of work has been reported about the design of suitable loss functions for medical image segmentation tasks.
\subsubsection{Cross Entropy Loss}
For image segmentation tasks, the cross entropy is one of the most popular loss functions. The function compares pixel-wisely the predicted category vector with the real segmentation result vector. For the case of binary segmentation, let $P(Y=1)=p$ and $P(Y=0)=1-p$, then the prediction is given by the sigmoid function, where $P(\hat{Y}=1)=1/(1+e^{-x})=\hat{p}$ and  $P(\hat{Y}=0)=1-1/(1+e^{-x})=1-\hat{p}$, $x$ is the output of neural networks. The cross entropy loss is defined as
\begin{equation}
\begin{aligned}
CE(p,\hat{p})=-(plog(\hat{p})+(1-p)log(1-\hat{p})).
\end{aligned}
\end{equation}
\subsubsection{Weighted Cross Entropy Loss}
The cross entropy loss deals with each pixel of images equally, and thus outputs an average value, which ignores the class imbalance and leads to a problem that the loss function depends on the class including the maximal number of pixels. Therefore, the cross entropy loss often shows low performance for small target segmentation.

To address the problem of class imbalance, Long et al. ~\cite{long2015fully} proposed weighted cross entropy loss (WCE) to counteract the class imbalance. For the case of binary segmentation, the weighted cross entropy loss is defined as
\begin{equation}
\begin{aligned}
WCE(p,\hat{p})=-(\beta plog(\hat{p})+(1-p)log(1-\hat{p})),
\end{aligned}
\end{equation}
where $\beta$ is used to tune the proportion of positive and negative samples, and it is an empirical value. If $\beta>1$, the number of false negatives will be decreased; on the contrary, the number of false positives will be decreased when $\beta<1$. In fact, the cross entropy is a special case of the weighted cross entropy when $\beta=1$. To adjust the weight of positive and negative samples simultaneously, we can use the balanced cross entropy (BCE) loss function that is defined as
\begin{equation}
\begin{aligned}
BCE(p,\hat{p})=-(\beta plog(\hat{p})+(1-\beta)(1-p)log(1-\hat{p})).
\end{aligned}
\end{equation}

In ~\cite{ronneberger2015u}, Ronneberger et al. proposed U-Net in which the cross entropy loss function is improved by adding a distance function. The improved loss function is able to improve the learning capability of models for inter-class distance. The distance function is defined as
\begin{equation}
\begin{aligned}
D(x)=\omega_0×e^{\frac{-(d_1(x)+d_2(x)^2}{2 \sigma^2}},
\end{aligned}
\end{equation}
where both $d_1(x)$ and $d_2(x)$ denote the distance between the pixel $x$ and boundaries of the first two nearest cells. So the final loss function is defined as
\begin{equation}
\begin{aligned}
L=\ BCE\left(p,\hat{p}\right)+\ D(x).
\end{aligned}
\end{equation}

\subsubsection{Dice Loss}
The Dice is a popular performance metric for the evaluation of medical image segmentation. This metric is essentially a measure of overlap between a segmentation result and corresponding ground truth. The value of Dice ranges from 0 to 1. “1” means the segmentation result completely overlaps with the real segmentation result. The calculation formula is defined as
\begin{equation}
\begin{aligned}
Dice\left(A,B\right)=\frac{2\times\left|A\cap B\right|}{A+B}\times100\%,
\end{aligned}
\end{equation}
where $A$ is a predicted segmentation result and $B$ is a real segmentation result.

For 3D medical volume data segmentation, Milletari et al. ~\cite{milletari2016v} proposed V-Net that employs the Dice loss
\begin{equation}
\begin{aligned}
DL(p,\hat{p})=1-\frac{2<p,\hat{p}>}{\left \| p \right\|_1+\left \| p \right\|_2},
\end{aligned}
\end{equation}
where $<p,\hat{p}>$ represents the dot product of the ground truth of each channel and the prediction result matrix.

It is worth noting that the Dice loss is suitable for uneven samples. However, the use of the Dice loss easily influences the back propagation and leads to a training difficulty. Besides, the Dice loss has a low robustness for different models such as mean surface distance or Hausdorff surface distance due to unbelievable gradient values. For example, the gradient of softmax function can be simplified to ($p-t$), where $t$ is the target value, and $p$ is the predicted value, but the value of dice loss is $2t^2$/${(p+t)}^2$. If values of $p$ and $t$ are too small, then the gradient value will change drastically leading to training difficulty.

\subsubsection{Tversky Loss}
Salehi et al. ~\cite{salehi2017tversky} proposed the Tversky Loss (TL) that is a regularized version of Dice loss to control the contribution of both false positive and false negative to the loss function. The TL is defined as
\begin{equation}
\begin{aligned}
TL(p,\hat{p})=\frac{p,\hat{p}}{p,\hat{p}+\beta(1-p,\hat{p})+(1-\beta)(p,1-\hat{p})},
\end{aligned}
\end{equation}
where $p \in {0,1}$ and $0 \leq \hat{p} \leq 1$. $p$ and $\hat{p}$ are the ground truth and predicted segmentation, respectively. TL is equivalent to (7) if $\beta=0.5$.

\subsubsection{Generalized Dice Loss}
Although the Dice loss can solve the problem of class imbalance to a certain extent, it does not work for serious class imbalance. For instance, small targets suffer from prediction errors of some pixels, which easily causes a large change for Dice values. Sudre et al. ~\cite{sudre2017generalised} proposed an Generalized Dice Loss (GDL), the GDL is defined as
\begin{equation}
\begin{aligned}
GDL\left(p,\hat{p}\right)=1-\frac{1}{m}\frac{2\sum_{j=1}^{m}{\omega_j\sum_{i=1}^{n}{p_{ij}{\hat{p}}_{ij}}}}{\sum_{j=1}^{m}{\omega_j\sum_{i=1}^{n}{(p_{ij}{+\hat{p}}_{ij})}}},
\end{aligned}
\end{equation}
where the weight $\omega=\left[\omega_1,\omega_2,...,\omega_m\right]$ is assigned to each class, and $\omega_j=1/{(\sum_{i=1}^{n}p_{ij})}^2$. The GDL is superior to the Dice loss since different areas have the similar contributions to the loss, and the GDL is more stable and robust during the training process.

\subsubsection{Boundary Loss}
To solve the problem of class imbalance, Kervadec et al. ~\cite{kervadec2019boundary} proposed a new boundary loss used for brain lesion segmentation. This loss function aims to minimize the distance between segmented boundaries and labeled boundaries. Authors conducted experiments on two imbalanced datasets with labels. The results show that the combination of the Dice loss and the boundary loss is superior to the single one. The composite loss is defined as
\begin{equation}
\begin{aligned}
L=\alpha L_{GD}(\theta)+(1-\alpha)L_B(\theta) ,
\end{aligned}
\end{equation}
where the first part is a regularized Dice Loss that is defined as
\begin{equation}
\begin{aligned}
L_{GD}\left(\theta\right)=1-2(\omega_G\sum_{p\epsilon\Omega}{g(p)s_\theta(p)} \\
+\omega_B\sum_{p\epsilon\Omega}{(1-g(p))(1-s_\theta(p))})/ \\
((\omega_G\sum_{p\epsilon\Omega}\left[g(p){+s}_\theta(p)\right] \\
+\omega_B\sum_{p\epsilon\Omega}{(2-g(p)-s_\theta(p))})),
\end{aligned}
\end{equation}
and the second part is the boundary loss that is defined as
\begin{equation}
\begin{aligned}
L_B\left(\theta\right)=\emptyset G(p)s_\theta(p),
\end{aligned}
\end{equation}
where if $p\epsilon G$, then $\emptyset G\left(p\right)=-||p-z_{\vartheta G}(p)||$, otherwise $\emptyset G\left(p\right)=||p-z_{\vartheta G}(p)||$. Besides, $\sum_{\mathrm{\Omega}}{g(p)f(s_\theta(p))}$ is used for the foreground and $\sum_{\mathrm{\Omega}}{(1-g(p))(1-f(s_\theta(p)))}$ is used for the background. The $L_{GD}\left(\theta\right)$ weight is $\omega_G=1/{(\sum_{p\epsilon\Omega}{g(p)})}^2$ and the $\omega_B=1/{(\sum_{p\epsilon\Omega}{(1-g(p))})}^2$. The $\mathrm{\Omega}$ represents the pixel set in the entire spatial domain.

\subsubsection{Exponential Logarithmic Loss}
In (9), the weighted dice loss is actually that the obtained dice value divides the sum of each label, which achieves a balance for objects with different scales. Therefore, by combining focal loss ~\cite{lin2017focal} and dice loss, Wong et al. ~\cite{wong20183d} proposed the exponential logarithmic loss (EXP loss) used for brain segmentation to solve problem of serious class imbalance. With the introduction of the exponential form, the nonlinearity of the loss functions can be further controlled to improve the segmentation accuracy. The EXP loss function is defined as
\begin{equation}
\begin{aligned}
L_{EXP}=\omega_{dice}\times L_{dice}+\omega_{cross}\times L_{cross},
\end{aligned}
\end{equation}
where two new parameter weights are denoted by $\omega_{dice}$ and $\omega_{cross}$, respectively. The $L_{dice}$ is an exponential log Dice loss, and the $L_{cross}$is a cross entropy loss
\begin{equation}
\begin{aligned}
L_{dice}=E[{(-ln({Dice}_i))}^{\gamma_{Dice}}],
\end{aligned}
\end{equation}
\begin{equation}
\begin{aligned}
L_{cross}=E[{\omega_l(-ln(p_l(x)))}^{\gamma_{cross}}],
\end{aligned}
\end{equation}
and,
\begin{equation}
\begin{aligned}
{Dice}_i=\frac{2(\sum_{x}{\sigma_{il}(x)p_i(x)})+\varepsilon}{\sum_{x}{{(\sigma}_{il}(x)+p_i(x)})+\varepsilon},
\end{aligned}
\end{equation}
\begin{equation}
\begin{aligned}
\omega_l={(\frac{\sum_{k} f_k}{f_l})}^{0.5},
\end{aligned}
\end{equation}
where $x$ is pixel position, $i$ is the label and $l$ is the ground-truth value at the position $x$. The $p_i(x)$ is the probability value outputted from the softmax.

In (17), $f_k$ is the frequency of occurrence of the label $k$, this parameter can reduce the influence of more frequently seen labels. Both $\gamma_{Dice}$ and $\gamma_{cross}$ are used to enhance the nonlinearity of the loss function.
\subsubsection{Loss Improvements}
For medical image segmentation, the improvement of loss mainly focuses on the problem of segmentation of small objects in a large background (the problem of class imbalance). Chen et al. ~\cite{chen2019learning} proposed a new loss function by applying traditional active contour energy minimization to convolutional neural networks, Li et al. ~\cite{li2020transformation} proposed a new regularization term to improve the cross-entropy loss function, and Karimi et al. ~\cite{karimi2019reducing} proposed a loss function based on Hausdorff distance (HD). Besides, there are still a lot of works ~\cite{taghanaki2019combo}~\cite{caliva2019distance} trying to deal with this problem by adding penalties to loss functions or changing the optimization strategy according to specific tasks.

In many medical image segmentation tasks, there are often only one or two targets in an image, and the pixel ratio of targets is sometimes small, which makes network training difficult. Therefore, to improve network training and segmentation accuracy, it is easier to focus on smaller targets by changing loss functions than to change the network structure. However, the design of loss functions is highly task-specific, so we need to analyze carefully task requirement, and then design reasonable and available loss functions.

\subsubsection{Deep supervision}
In general, the increase of network depth can improve the feature representation of networks to some extent, but it simultaneously causes new problems such as vanishing gradient and gradient explosion. In order to train deep networks effectively, Lee et al. ~\cite{lee2015deeply} proposed Deeply-supervised nets (DSNs) by adding some auxiliary branching classifiers to some layers of the neural network. Dou et al. ~\cite{dou20173d} proposed a 3D DSN for heart and liver segmentation, which incorporates a 3D deep monitoring mechanism into a 3D full convolutional network for volume-to-volume learning and inference, eliminating redundant computation and reducing the risk of over-fitting in the case of limited training data. Similarly, Dou et al ~\cite{dou2020deep} presented a method for fetal brain MRI cortical plate segmentation using a fully convolutional neural network architecture with deep supervision and residual connection, and obtained high segmentation accuracy for brain MRI cortical plate segmentation. In fact, deep supervision not only can constrain the discrimination and robustness of learned features at all stages, but also improves network training efficiency.

\section{Weakly supervised learning}
Although convolutional neural networks show strong adaptability for medical image segmentation, segmentation results seriously depend on high-quality labels. In fact, it is rare to build many datasets with high-quality labels, especially in the field of medical image analysis, since data acquisition and labeling often incur high costs. Therefore, a lot of studies on incomplete or imperfect datasets are reported. We summarize these studies as weakly supervised learning as shown in Fig. 14.

\begin{figure}[htbp]
	\centerline{\includegraphics[width=\columnwidth]{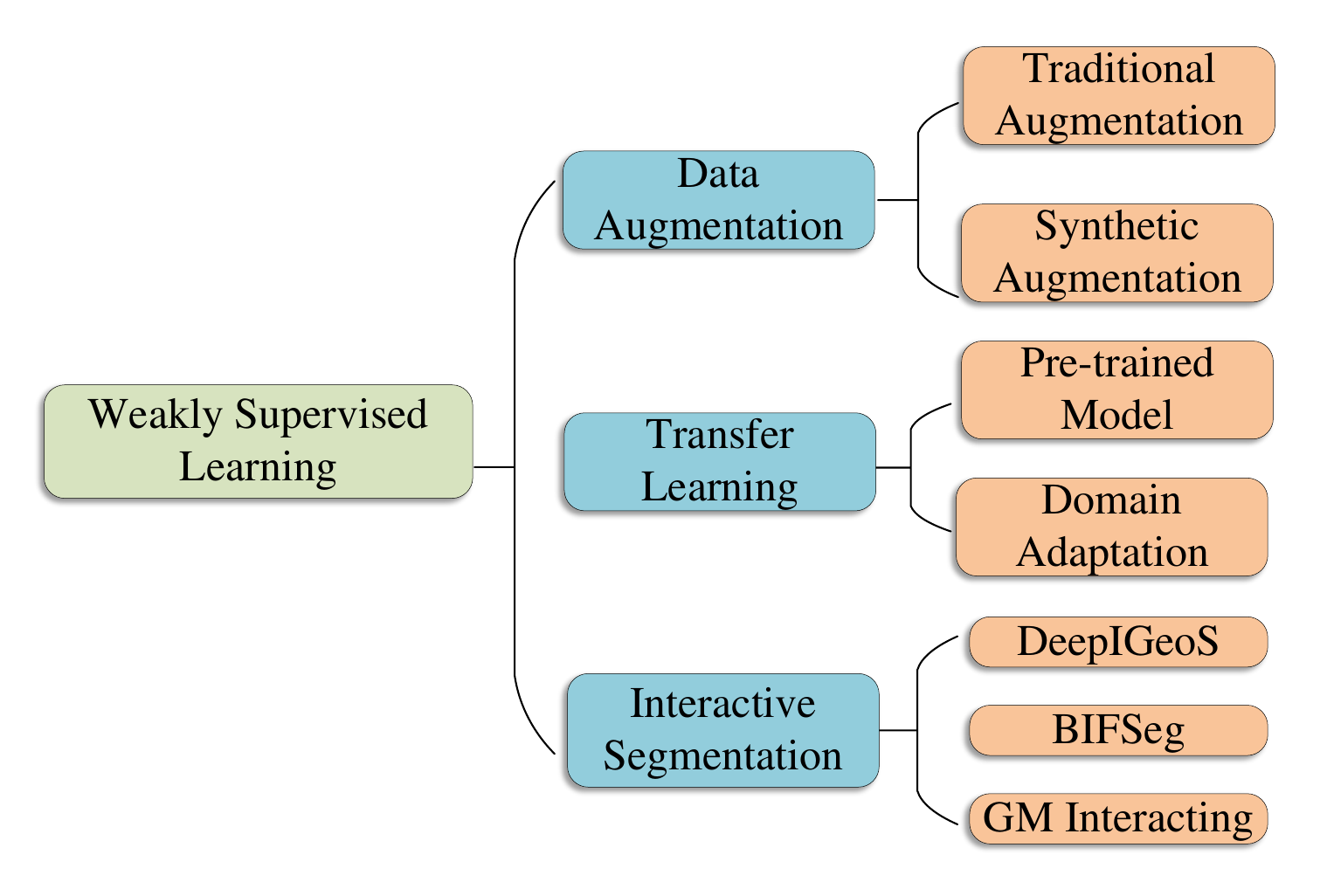}}
	\caption{The weakly supervised learning methods for medical image segmentation.}
	\label{fig14}
\end{figure}

\subsection{Data Augmentation}
In the absence of largely labeled datasets, data augmentation is an effective solution to this problem. However, general data expansion methods produce images that are highly correlated with original images. Compared to common data augmentation approaches, GAN proposed by Goodfellow ~\cite{goodfellow2014generative} is currently a popular strategy for data augmentation since GAN overcomes the problem of reliance on original data.

\emph{Traditional Methods:} General data augmentation methods include the improvement of image quality such as noise suppression, the change of image intensity such as brightness, saturation, and contrast, and the change of image layout such as rotation, distortion, and scaling, etc. Sirinukunwattana et al. ~\cite{sirinukunwattana2017gland} utilized the Gaussian blur to achieve data enhancement, which is helpful for performing gland segmentation tasks in the colon tissue images. Dong et al. ~\cite{dong2017automatic} randomly used the brightness enhancement function in 3D MR images to enrich training data for brain tumor segmentation. Contrast enhancement is usually helpful when an image shows uneven intensity. Furthermore, Ronneberger et al. ~\cite{ronneberger2015u} used random elastic deformation to perform data expansion on the original dataset. In fact, the most commonly method used for traditional data augmentation is parametric transformation (rotation, translation, shear, shift, flip, ...). Since this kind of transformation is virtual without computational cost and the annotation on medical images is difficult, it is always performed before each training session.

\emph{Conditional Generative Adversarial Nets (cGAN):} In contrast to the use of cGAN for supervised learning introduced in Section \uppercase\expandafter{\romannumeral2}, this section focuses on the use of cGAN for data augmentation. An original GAN generator denoted by $G$ can learn data distribution, but generated pictures are random, which means that the generation process of the $G$ is an unguided state. In contrast, cGAN adds a condition to the original GAN in order to guide the generation process of the $G$. Fig. 15 shows the architecture of cGAN. Guibas et al. ~\cite{guibas2017synthetic} proposed a network architecture composed of a GAN ~\cite{goodfellow2014generative} and a cGAN ~\cite{mirza2014conditional}. The random variables are input into the GAN leading to the generation of a synthetic image of fundus blood vessel label, then the generated label map is input into the conditional GAN to generate a real retinal fundi image. Finally, authors verified the authenticity of synthesized images by checking whether the classifier can distinguish a synthesized image from a real image. Mahapatra et al. ~\cite{mahapatra2018efficient} used a cGAN to synthesize X-ray images with required abnormalities, this model considers abnormal X-ray images and lung segmentation labels as inputs, and then generates synthetic X-ray images with same diseases as input X-ray images. At the same time, the segmentated label is obtained. In addition, there are also some other works ~\cite{shin2018medical}~\cite{jin2018ct} using GAN or cGAN to generate images to achieve data enhancement. Although the image generated by cGAN has many defects, such as blurred boundary and low resolution, the cGAN provides a basic ideas for the later CycleGAN ~\cite{zhu2017unpaired} and StarGAN ~\cite{choi2018stargan} used for the conversion of image styles.

\begin{figure}[htbp]
	\centerline{\includegraphics[width=\columnwidth]{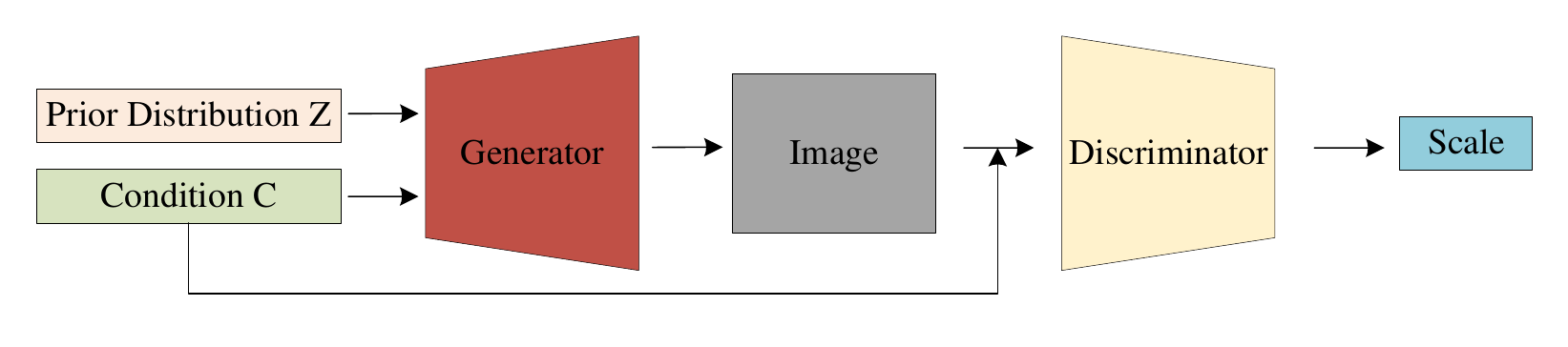}}
	\caption{The cGAN architecture ~\cite{mirza2014conditional}.}
	\label{fig15}
\end{figure}

\subsection{Transfer Learning}
By utilizing trained parameters of a model to initialize a new model, transfer learning can achieve fast model training for data with limited labels. One approach is to fine-tune the pre-trained model on ImageNet for the target medical image analysis task, while the other is to migrate the training for data from across domains.

\emph{Pre-trained Model:} Transfer learning is often used to solve the problem of limited data labeled in medical image analysis, and some researchers found that using pre-trained networks on natural images such as ImageNet as an encoder within a U-Net-like network and then performing fine-tuning on medical data can further improve the segmentation effect of medical images. Kalinin ~\cite{kalinin2020medical} et al. considered the VGG-11, VGG-16, and ResNet-34 networks pre-trained on ImageNet as encoders of the U-shaped network to perform semantic segmentation of robotic instruments from wireless capsule endoscopic videos of vascular proliferative lesions and surgical procedures. Similarly, Conze et al. ~\cite{conze2020healthy} used VGG-11 pre-trained on ImageNet as the encoder of a segmentation network to perform shoulder muscle MRI segmentation. Experiments demonstrate that the pre-trained network is useful for improving segmentation accuracy. It can be concluded that a pre-trained model on ImageNet can learn some common underlying features that are required for both medical and natural images, thus retraining process is unnecessary while performing fine-tuning is useful for training models. However, the domain adaptive may be a problem when applying pre-trained models of natural scene images to medical image analysis tasks. Besides, popular transfer learning methods are hardly applicable to 3D medical image analysis because pre-trained models often rely on 2D image datasets. If the number of medical datasets with annotations is large enough, it is possible that the effect of pre-training is weak for improving model performance. In fact, the effect of a pre-trained model is unstable and it depends on segmentation datasets and tasks. Empirically, we can try to use the pre-trained model if it can improve segmentation accuracy, otherwise we need to consider designing new models.

\emph{Domain Adaptation:}If the labels from the training target domain are not available, and we can only access the labels in other domains, then popular methods are to transfer the trained classifier on the source domain to the target domain without labeled data. CycleGAN is a cycle structure, and mainly composed of two generators and two discriminators. Fig. 16 shows the architecture of CycleGAN. First, an image in the X domain is transferred to the Y domain by a generator G, and then the output from the G is reconstructed back to the original image in the X domain by the generator F. On the contrary, the image in the Y domain is transferred to the X domain by the generator F, and then the output from the F is reconstructed back to the original image in the Y domain by the generator G. Both discriminator G and F play discriminating roles ensuring the style transfer of images. Huo et al. ~\cite{huo2018adversarial} proposed a jointly optimized image synthesis and segmentation framework for the task of spleen segmentation in CT images using CycleGAN ~\cite{zhu2017unpaired}. The framework achieves an image conversion from the marked source domain to the synthesized target domain. During training, synthesized target images are used to train the segmentation network. During the test process, a real image from the target domain is directly input into the trained segmentation network to obtain desired segmentation results. Chen et al. ~\cite{chen2019synergistic} also adopted a similar method using segmentation labels of MR images to achieve the task of cardiac CT segmentation.

\begin{figure}[htbp]
	\centerline{\includegraphics[width=\columnwidth]{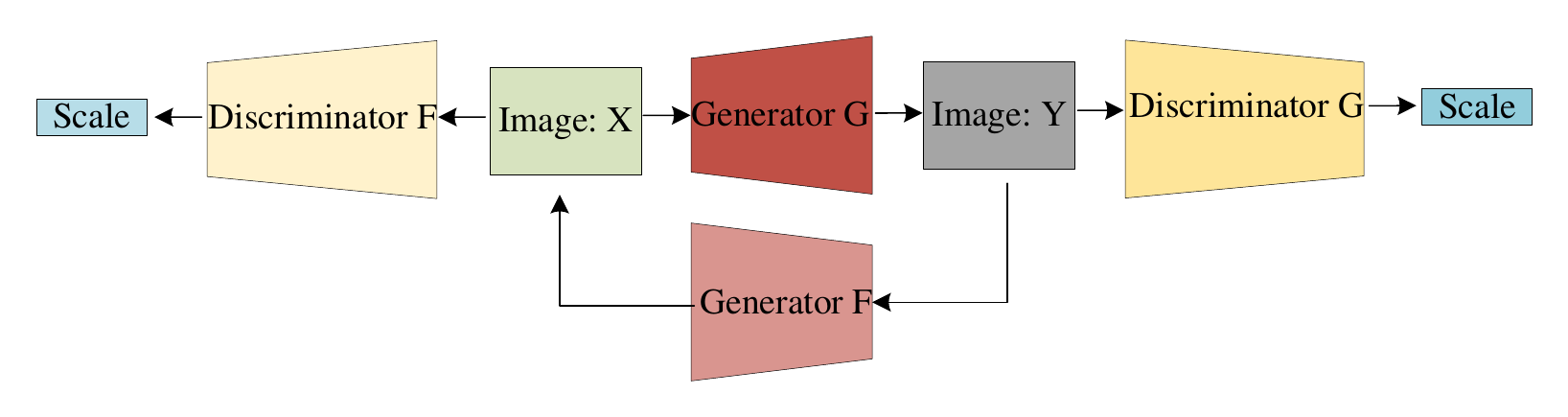}}
	\caption{The Cycle GAN architecture ~\cite{zhu2017unpaired}.}
	\label{fig16}
\end{figure}

Chartsias et al. ~\cite{chartsias2017adversarial} used the CycleGAN to generate corresponding MR images and labels from CT slices and myocardial segmentation labels, and then used synthetic MR and real MR images to train the myocardial segmentation model. This model obtains 15\% improvement over the myocardial segmentation model trained on real MR images. Similarly, there are some other works  that realize the image conversion between different domains through the CycleGAN and improve the performance of medical image segmentation ~\cite{zhao2017whole}~\cite{valindria2018multi}.

\subsection{Interactive Segmentation}
Manually drawing medical image segmentation labels is usually tedious and time-consuming, especially for the drawing of 3D volume data. Interactive segmentation allows clinicians to correct interactively the initially segmented image generated by a model to obtain more accurate segmentation. The key to effective interactive segmentations is that clinicians can use interactive methods such as mouse clicks and outline boxes to improve an initial segmentation result from a model. Then the model can update parameters and generate new segmentation images to obtain new feedback from the clinicians.

Wang et al. ~\cite{wang2018deepigeos} proposed the DeepIGeoS using the cascade of two CNNs for interactive segmentation of 2D and 3D medical images. The first CNN called P-Net outputs a coarse segmentation result. Based on this, users provide interactive points or short lines to mark wrong segmentation areas, and then use them as the input of the second CNN called R-Net to obtain corrected results. Experiments were conducted on two dimensional foetal MRI images and three-dimensional brain tumor images, and experimental results showed that compared with traditional interactive segmentation methods such as GraphCuts, RandomWalks and ITK-Snap, the DeepIGeoS greatly reduces the requirement for user interaction and reduces user time.

Wang et al. ~\cite{wang2018interactive} proposed the BIFSeg that is similar to the principle of GrabCut ~\cite{boykov2001interactive}~\cite{rother2004grabcut}. Users first draw a bounding box, and the area inside the bounding box is considered as the input of CNN, then an initial result is obtained. After that, users perform an image-specific fine-tuning to make CNN provide better segmentation results. The GrabCut achieves image segmentation by learning a Gaussian mixture model (GMM) from images, while the BIFSeg learns a CNN from images. Usually CNN-based segmentation methods can only deal with objects that have appeared in the training set, which limits the flexibility of these methods, but the BIFSeg attempts to use a CNN to segment objects that have not been seen during training process. The process is equivalent to making the BIFSeg learn to extract the foreground part of the object from a bounding box. During the test, the CNN can better use the information in the specific image through an adaptive fine-tuning.

Rupprecht et al. ~\cite{rupprecht2018guide} proposed a new interactive segmentation method named GM interacting that updates image segmentation results according to the input text from users. This method changes the output of the network by modifying the feature maps between an encoder and a decoder interactively. The category of areas is first set according to the response of users, then some guiding parameters including multiplication and offset coefficients are updated through back propagation, the feature map is finally changed resulting in updated segmentation results.

The interactive image segmentation based on deep learning can reduce the number of user interactions and the user time, which shows broader application prospects.

\subsection{Others Works}
Semi-supervised learning can use a small part of labeled data and any number of unlabeled data to train a model, and its loss function often consists of the sum of two loss functions. The first is a supervised loss function that is only related with labeled data. The second is an unsupervised loss function or regularization term that is related to both labeled and unlabeled data.

Based on the idea of GAN, Zhang et al. ~\cite{zhang2018translating} proposed a semi-supervised learning framework based on the adversarial way between segmentation network and evaluation network. An image is fed into U-Net to generate a segmentation map, which is then stacked with the original image and presented to the evaluation network to obtain a segmentation score. During the training process, the segmentation network is optimized in two aspects, one is to minimize the segmentation loss of labeled images and the other is to make the evaluation network obtain high scores for unlabeled images. Besides, the evaluation network is updated to assign low scores to unmarked images but high scores to marked images. Due to this adversarial learning, the segmentation network obtains supervised signals from both labeled and unlabeled images. Thus, the semi-supervised learning framework achieves better segmentation effect in the gland segmentation task for histopathology images. Similarly, some other semi-supervised frameworks ~\cite{baur2017semi}~\cite{chartsias2018factorised}~\cite{li2020transformation}~\cite{zhao2019data} are also proposed to optimize medical image segmentation.

Accurate and robust segmentation of organs or lesions from medical images plays a vital role in many clinical applications, such as diagnosis and treatment planning. However, it is difficult for medical images to acquire the annotated data, as generating accurate annotations requires expertise and time. Weakly supervised segmentation methods learn image segmentation from border or image-level labels or from a small amount of annotated image data, rather than using a large number of pixel-level annotations, to obtain high-quality segmentation results. In fact, a small amount of annotated data and a large amount of unannotated data are more compatible with the real clinical situation. However, in practice, the performance of weakly supervised learning only provides rarely acceptable results for medical image segmentation tasks, especially for 3D medical images. Therefore, this is a direction worth exploring in the future.

\section{Currently popular direction}
\subsection{Network Architecture Search}
Recently, the performance of convolutional neural network models has been continuously improved. Researchers have designed a large number of popular network architectures for specific tasks such as image classification, segmentation, reconstruction, etc. These architectures are often designed by industry experts or academics for months or even years, since the design of network architectures with excellent performance usually requires a great deal of domain knowledge. Therefore, the design process is time consuming and laborious for researchers without domain knowledge. So far, NAS ~\cite{elsken2018neural}has made significant progress in improving the accuracy of image classification. The NAS can be deemed to a subdomain of automatic machine learning ~\cite{he2020automl}(AutoML) and has a strong overlap with hyperparametric optimization ~\cite{ha2019bayesian}and meta learning ~\cite{vanschoren2018meta}. Current research on NAS focuses on three aspects: search space, search strategy and performance estimation. The search space is a candidate collection of network structures to be searched. The search space is divided into a global search space that represents the search for the entire network structure, and a cell-based search space that searches only a few small structures that are assembled into a complete large network by the ways of stacking and stitching. The search strategy aims to find the optimal network structure as fast as possible in search spaces. Popular search strategies are often grouped into three categories, reinforcement-based learning, evolutionary algorithms, and gradients. Performance estimation strategy is the process of assessing how well the network structure performs on target datasets. For NAS techniques, researcher pay more attention to the improvement of search strategies since search space and performance estimation methods are usually rarely changed. Some improved CNN model based on NAS ~\cite{chen2018searching}~\cite{liu2019auto}have been proposed and applied to image segmentation.

Most current studies on deep learning in medical image segmentation depend on U-Net networks and makes some changes to the network structure according to different tasks, but in reality the non-network structure factors may be also important for improving segmentation effect. Isensee et al.~\cite{isensee2018nnu} argued that too much manual adjustment on network structure could lead to over-fitting for a given dataset, and therefore proposed a medical image segmentation framework no-new-UNet (nnU-Net) that adapts itself to any new dataset. The nnUnet automatically adjusts all hyperparameters according to the properties of the given dataset without manual intervention. Therefore, the nnU-Net only relies on vanilla 2D UNet, 3D UNet, UNet cascade and a robust training scheme. It focuses on the stage of pre-processing (resampling and normalization), training (loss, optimizer settings, data augmentation), inference (patch-based strategies, test-time-augmentations integration, model integration, etc.), and post-processing (e.g., enhanced single pass domain). In practical applications, the improvements of network structure design usually depend on experiences without adequate interpretability theory support, Moreover, more complex network models indicate higher risk of over-fitting.

Weng et al ~\cite{weng2019unet}first proposed a NAS-UNet for medical image segmentation. The NAS-UNet contains the same two cell architectures DownSC and UpSC. The difference between them is that the former performs a search on the U-shaped backbone to obtain DownSC and UpSC blocks. The NAS-UNet outperforms the U-Net and its variants, and its training time is close to that of U-Net, but with only 6\% of the number of parameters.

To perform image segmentation in real time for high-resolution 2D images (e.g. CT, MRI and histopathology images), the study of compressed neural network models has become a popular direction in medical image segmentation. The application of NAS can effectively reduce the number of model parameters and achieves high segmentation performance. Although the performance of NAS is stunning,the fact of why particular architectures perform well can not be explained. Therefore, it is also important for future research to better understand the mechanisms which have a significant impact on performance and to explore whether these properties can be generalized to different tasks.

\subsection{Graph Convolutional Neural Network}
The GCN ~\cite{wu2020comprehensive}is one of the powerful tools for the study of non-Euclidean domains. A graph is a data structure consisting of nodes and edges. The early graph neural networks (GNNs) ~\cite{scarselli2008graph} mainly address strictly graphical problems such as the classification of molecular structures. In practice, the Euclidean spaces (e.g., images) or sequences (e.g., text), and many common scenes can be converted into graphs that can be modeled by using GCN techniques.

Gao et al. ~\cite{gao2019graph} designed a new graph pooling (gPool) and graph unpooling (gUnpool) operation based on GCN and proposed an encoder-decoder model namely graph U-Net. The graph U-Net achieves better performance than popular U-Nets by adding a small number of parameters. In contrast to traditional convolutional neural networks where deeper is better, the performance of the graph U-Net cannot be improved by increasing the depth of networks when the value of depth exceeds 4. However, the graph U-Net show stronger capability of feature encoding than popular U-Nets when the value of depth is smaller or equivalent to 4. Yang et al. ~\cite{yang2020cpr} proposed the end-to-end conditional partial residual plot convolutional network CPR-GCN for automatic anatomical marking of coronary arteries. Authors showed that the GCN-based approach provided better performance and stronger robustness than traditional and recent depth learning based approaches. Results from these GCNs in medical image segmentations are promising, as the graph structure has high data representation efficiency and strong capability of feature encoding.

\subsection{Interpretable Shape Attentive Neural Network}
Currently, many deep learning algorithms tend to make judgments by using "memorized" models that approximately fit to input data. As a result, these algorithms cannot be explained sufficiently and give convincing evidences for each specific prediction. Therefore, the study of the interpretability of deep neural networks is a hot topic at present.

Sun et al. ~\cite{sun2020saunet} proposed the SAU-Net that focuses on the interpretability and the robustness of models. The proposed architecture attempts to address the problem of poor edge segmentation accuracy in medical images by using a secondary shape stream. Specially, the shape stream and the regular texture stream can capture rich shape-dependent information in parallel. Furthermore, both spatial and channel attention mechanism are used for the decoder to explain the learning capability of models at each resolution of U-Net. Finally, by extracting the learned shape and spatial attention maps, we can interpret the highly activated regions of each decoder block. The learned shape maps can be used to infer correct shapes of interesting categories learned by the model. The SAU-Net is able to learn robust shape features of objects via the gated shape stream, and is also more interpretable than previous works via built-in saliency maps using attention.

Wickstrøm et al. ~\cite{wickstrom2020uncertainty} explored the uncertainty and interpretability of semantic segmentation of colorectal polyps in convolutional neural networks, and the authors developed the central idea of guided back propagation ~\cite{springenberg2014striving} for the interpretation of network gradients. By using back propagation, the gradient corresponding to each pixel in the input is obtained so that the features considered by the network can be visualized. In the process of back propagation, pixels with large and positive gradient values in an image should be paid more attention due to high importance while pixels with large and negative gradient values should be suppressed. If these negative gradients are included in the visualization of important pixels, they may result in noisy visualizations of descriptive features. To avoid creating noisy visualizations, the guide back propagation process changes the back propagation of the neural network so that the negative gradients are set to zero at each layer, thereby allowing only positive gradients to flow backwards through the network and highlight these pixels.

Medical image analysis is an aid to the clinical diagnosis, the clinicians wonder not only where the lesion is located at, but also the interpretability of results given by networks. Currently, the interpretation of medical image analysis is dominated by visualization methods such as attention and the class-activation-map (CAM). Therefore, the research on the interpretability of deep learning for medical image segmentation ~\cite{guan2018diagnose}~\cite{tang2019interpretable}~\cite{selvaraju2017grad}~\cite{zhang2019pathologist} will be a popular direction in future.

\subsection{Multi-modality Data Fusion}
Multi-modality data fusion has been widely used in medical image analysis because it can provide richer object features that are helpful for improving object detection and segmentation results. Dou et al. ~\cite{dou2020unpaired} proposed a novel multi-modal learning scheme for accurate segmentation of anatomical structures from unpaired CT and MRI images, and designed a new loss function using knowledge distillation to improve model training efficiency ~\cite{hinton2015distilling}. More specifically, the normalization layer used for different modalities (i.e., CT and MRI) is implemented within separate variables, whereas the convolutional layer is constructed within shared variables. In each training iteration, samples for each modality are loaded separately and then forwarded to the shared convolutional and independent normalization layers, and finally the logarithms that can be used to calculate knowledge distillation losses will be obtained. Moeskops et al. ~\cite{moeskops2016deep} investigated a question whether it is possible to train a single convolutional neural network (CNN) to perform same segmentation tasks on different-modality data. It is well known that CNNs show excellent performance for image feature encoding and based on this, the experiments in ~\cite{moeskops2016deep} furthermore demonstrate that CNNs are also excellent for feature encoding of multi-modality data when they are used for the same tasks. Therefore, a single system can be used in clinical practice to automatically execute segmentation tasks on various modality data without extra task-specific training.

More relevant literatures can be found in the review on multi-modal fusion for medical image segmentation using deep learning ~\cite{zhou2019review}. In this review, authors classified fusion strategies into three categories: input-level fusion, layer-level fusion, and decision-level fusion. Although it is known that multi-modal fusion networks usually show better performance for segmentation tasks than unimodal networks, multi-model fusion causes some new problems such as how to design multi-modal networks to efficiently combine different modalities, how to exploit potential relationships between different modalities, how to integrate multiple information into segmentation networks to improve segmentation performance, etc. In addition, the integration of multi-modal data fusion into an effective single-parameter network can help simplify deployment and improve the usability of models in clinical practice.

\section{Discussion and outlook}
\subsection{Medical Image Segmentation Datasets}
In order to help clinicians make accurate diagnoses, it is necessary to segment important organs, tissues or lesions from medical images with the aid of a computer and extract features from segmented objects. As a result, various medical image datasets and corresponding competitions have been launched to promote the development of computer-aided diagnosis techniques. In recent years, there has been a growing interest in developing more comprehensive computational anatomical models with the development of deep learning techniques, which has facilitated the development of multi-organ analysis models. The multi-organ segmentation approaches are different from traditional organ-specific strategies in that they incorporate relationships between different organs into models to represent more accurately the complex human anatomy. In the context of multiorgan analysis, brain and abdomen are the most popular in medical image analysis. Thus there are many datasets on the brain and abdomen such as BRATS ~\cite{menze2014multimodal}~\cite{bakas2017advancing}~\cite{bakas2018identifying}, ISLES ~\cite{maier2017isles}, KITS ~\cite{heller2019kits19}, LITS ~\cite{bilic2019liver}, CHAOS ~\cite{kavur2019chaos}, etc. There are two reasons for the emergence of large datasets: on the one hand, the rapid development of imaging techniques, increasingly higher resolution shows more detailed anatomical tissue, which provides a better reference for clinicians; on the other hand, with the development of deep learning techniques, a large number of training samples are necessary, so many research teams have collected many samples and annotated data to form datasets in order to train network models easily. In addition, stable organ structures in the abdomen (e.g., the liver, spleen, and kidneys) can provide constraints and contextual information for creating computational anatomical models of the abdomen. There are also a small number of public datasets on hippocampus and pelvic organs (e.g., Colon ~\cite{sunrui2019rr}, and prostate ~\cite{litjens2014evaluation}). Indeed, the construction of more holistic and global anatomical models remains one of the greatest challenges and opportunities in future due to the lack of large datasets to characterize the complexity of the human anatomy. More discussions on multi-organ analysis and computational anatomical methods can be found in ~\cite{cerrolaza2019computational}. The review proposed by Cerrolaza et al. ~\cite{cerrolaza2019computational} follows a methodology-based classification of different techniques that are available for the analysis of multi-organs and multi-anatomical structures, from techniques using point distribution models to the latest deep learning-based approaches.

\begin{figure*}[htbp]
	\centerline{\includegraphics[width=\textwidth]{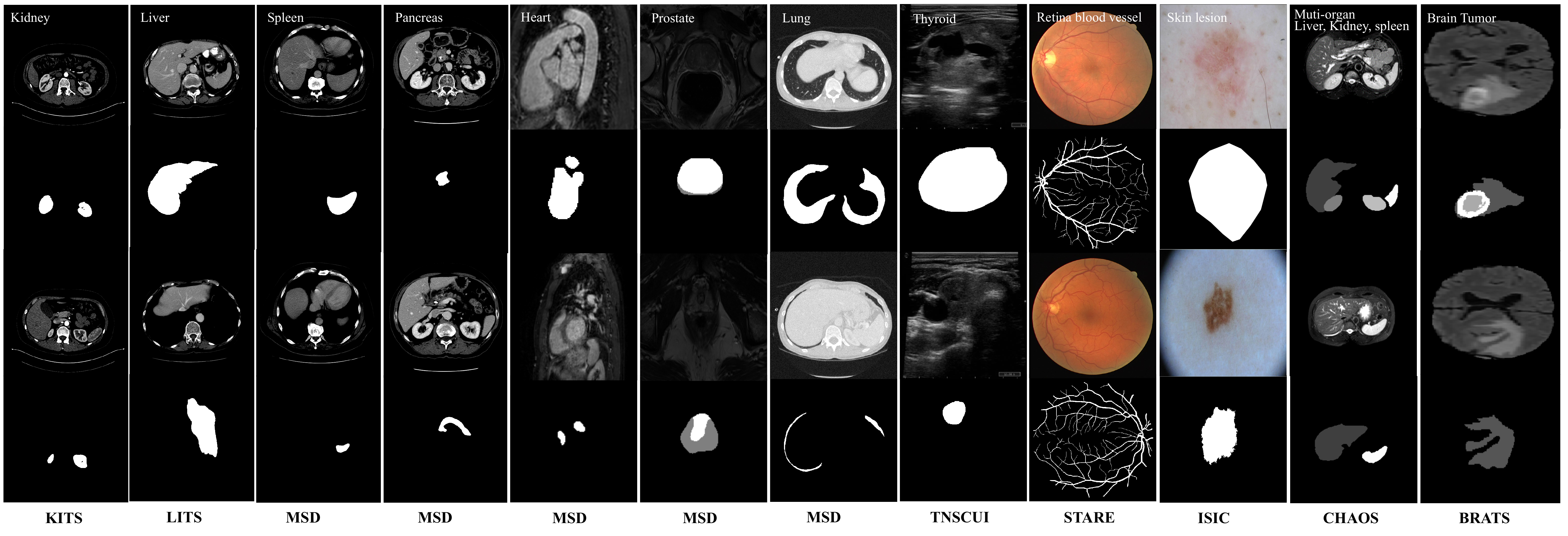}}
	\caption{Some images of benchmark datasets.}
	\label{fig1}
\end{figure*}

\begin{table*}[ht]
\caption{Public datasets for medical segmentation.}
	\setlength{\footnotesize\tabcolsep}{5pt}
\renewcommand\arraystretch{1.4}
    \resizebox{\textwidth}{!}{
\begin{tabular}{ l l l }
\hline
 \textbf{Objects}   & \textbf{Dataset}             & \textbf{URL}                                                                                           \\ \hline
\multicolumn{1}{l}{\multirow{4}{*}{Liver}}                                             & LiTS ~\cite{bilic2019liver}                      & https://competitions.codalab.org/competitions/17094                                                    \\ \hhline{~--}
\multicolumn{1}{c}{}                              & Sliver07 ~\cite{sunrui2019liver}           	 & http://www.sliver07.org/                                                                               \\  \hhline{~--}
\multicolumn{1}{c}{}                              & 3Dircadb ~\cite{france20163dircadb}       	 & https://www.ircad.fr/research/3dircadb/                                                                \\ \hhline{~--}
\multicolumn{1}{c}{}                              & Medical Segmentation Decathlon (MSD) ~\cite{simpson2019large}  & http://medicaldecathlon.com/index.html       \\ \hhline{~--}
\multicolumn{1}{c}{}                              &CHAOS ~\cite{kavur2020chaos}                  & https://chaos.grand-challenge.org                        \\  \hline
\multicolumn{1}{l}{\multirow{2}{*}{Pancreas}}    & Medical Segmentation Decathlon (MSD) ~\cite{simpson2019large}  & http://medicaldecathlon.com/index.html    \\  \hhline{~--}
\multicolumn{1}{c}{}	                           &  NIH Pancreas ~\cite{sunrui2019ss}          	 & http://academictorrents.com/details/80ecfefcabede760cdbdf63e38986501f7becd49                           \\ \hline
\multicolumn{1}{l}{\multirow{2}{*}{Colon}}                                         &  COLONOGRAPHY ~\cite{sunrui2019rr}     	 	 & https://wiki.cancerimagingarchive.net/display/Public/CT+COLONOGRAPHY\#dc149b9170f54aa29e88f1119e25ba3e \\ \hhline{~--}
\multicolumn{1}{c}{}                                            &  Medical Segmentation Decathlon (MSD) ~\cite{simpson2019large}  & http://medicaldecathlon.com/index.html    \\ \hline
\multicolumn{1}{l}{\multirow{2}{*}{Heart}}                                             &    AMRG Cardiac Atlas ~\cite{suinesiaputra2014big} 		 & http://www.cardiacatlas.org/studies/amrg-cardiac-atlas/                                                \\ \hhline{~--}
\multicolumn{1}{c}{}                                           &  Medical Segmentation Decathlon (MSD) ~\cite{simpson2019large}  & http://medicaldecathlon.com/index.html    \\ \hline
\multicolumn{1}{l}{\multirow{3}{*}{Lung}}                                               &   LIDC-IDRI ~\cite{armato2011lung}           	 & https://wiki.cancerimagingarchive.net/display/Public/LIDC-IDRI\#                                       \\ \hhline{~--}
\multicolumn{1}{c}{}                                              &  VESSEL12 ~\cite{topkaya2014counting}           	 & https://vessel12.grand-challenge.org/                                       \\ \hhline{~--}
\multicolumn{1}{c}{}                                             &  Medical Segmentation Decathlon (MSD) ~\cite{simpson2019large}  & http://medicaldecathlon.com/index.html    \\ \hline
\multicolumn{1}{l}{\multirow{2}{*}{Prostate}}                                          &  PROMISE12 ~\cite{litjens2014evaluation}   	 & https://promise12.grand‐challenge.org/                                                                 \\ \hhline{~--}
\multicolumn{1}{c}{}                                             &  Medical Segmentation Decathlon (MSD)~\cite{simpson2019large}  & http://medicaldecathlon.com/index.html    \\ \hline
\multicolumn{1}{l}{\multirow{4}{*}{Brain}}          & OASIS ~\cite{lamontagne2019oasis}      	 & http://www.oasis-brains.org/                                                                           \\ \hhline{~--}
\multicolumn{1}{c}{}                                &     BRATS ~\cite{menze2014multimodal}~\cite{bakas2017advancing}~\cite{bakas2018identifying}  & https://www.med.upenn.edu/sbia/brats2018/registration.html                                             \\ \hhline{~--}
\multicolumn{1}{c}{}                                 &  ISLES ~\cite{maier2017isles}                & http://www.isles‐challenge.org/                                                                        \\ \hhline{~--}
\multicolumn{1}{c}{}                                 &  mTOP ~\cite{kistler2013virtual}             & https://www.smir.ch/MTOP/Start2016                                                                     \\ \hhline{~--}
\multicolumn{1}{c}{}                                       &  Medical Segmentation Decathlon (MSD)~\cite{simpson2019large}  & http://medicaldecathlon.com/index.html    \\ \hline
\multicolumn{1}{l}{\multirow{2}{*}{Kidney}}                                            & KITS ~\cite{heller2019kits19}              & https://kits19.grand-challenge.org                                                                     \\ \hhline{~--}
\multicolumn{1}{c}{}                                            &CHAOS ~\cite{kavur2020chaos}                  & https://chaos.grand-challenge.org                        \\  \hline
\multicolumn{1}{l}{\multirow{2}{*}{Spleen}} 												 &  Medical Segmentation Decathlon (MSD)~\cite{simpson2019large}  & http://medicaldecathlon.com/index.html    \\ \hhline{~--}
\multicolumn{1}{c}{}   												& CHAOS ~\cite{kavur2020chaos}                  & https://chaos.grand-challenge.org                        \\  \hline
Hippocampus										     &  Medical Segmentation Decathlon (MSD)~\cite{simpson2019large}  & http://medicaldecathlon.com/index.html    \\ \hline
Hepatic Vessel										  &  Medical Segmentation Decathlon (MSD)~\cite{simpson2019large}  & http://medicaldecathlon.com/index.html    \\ \hline
Skin lesion                     &ISIC ~\cite{codella2018skin}    & https://challenge.isic-archive.com/data                            \\ \hline
STARE                         &STARE~\cite{hoover2000locating}    &https://cecas.clemson.edu/~ahoover/stare/  \\ \hline
Thyroid                         &TNSCUI~\cite{zhang2020cascade}    &https://tn-scui2020.grand-challenge.org/  \\ \hline
\end{tabular}}
\end{table*}

There are many publicly available datasets for medical image segmentation, Table I provides a brief description and list of each dataset. As shown in Fig. 17, we also provide some images of benchmark datasets. In fact, there are more public datasets than in the list of Table I used for medical image segmentation.

\subsection{Popular evaluation metrics}
In order to measure effectively the performance of medical image segmentation model, a large number of metrics have been proposed for evaluating the segmentation effectiveness. The evaluation of image segmentation performance relies on pixel quality, region quality and surface distance quality. In this section, we give some popular metrics for evaluating the performance of medical image segmentation. Pixel quality metrics include pixel accuracy (PA). Region quality metrics include Dice score, volume overlap error (VOE) and relative volume difference (RVD). Surface distance quality metrics include average symmetric surface distance (ASD) and maximum symmetric surface distance (MSD).

\textbf{\emph{PA:}} Pixel accuracy simply finds the ratio of pixels properly classified, divided by the total number of pixels. For $K+1$ classes ($K$ foreground classes and the background) pixel accuracy is defined as:
\begin{equation}
\begin{aligned}
PA=\frac{\sum_{i=0}^{K}p_{ii}}{\sum_{i=0}^{K}\sum_{j=0}^{K}p_{ij}},
\end{aligned}
\end{equation}
where $p_{ij}$ is the number of pixels of class $i$ predicted as belonging to class $j$.

\textbf{\emph{Dice score:}} it is a popular metric for image segmentation (and is more commonly used in medical image analysis), which can be defined as twice the overlap area of predicted and ground-truth maps, divided by the total number of pixels in both images. The Dice score is defined as:
\begin{equation}
\begin{aligned}
Dice=\frac{2\left|A\cap B\right|}{\left|A\right|+\left|B\right|},
\end{aligned}
\end{equation}
where $A$ and $B$ denote the ground truth and the predicted segmentation maps, respectively.

\textbf{\emph{VOE:}} it is the complement of the Jaccard index, it is defined as:
\begin{equation}
\begin{aligned}
VOE\left(A,B\right)=1-\frac{\left|A\cap B\right|}{\left|A\cup B\right|}.
\end{aligned}
\end{equation}

\textbf{\emph{RVD:}} it is an asymmetric measure defined as:
\begin{equation}
\begin{aligned}
RVD\left(A,B\right)=\frac{\left|B\right|-\left|A\right|}{\left|A\right|}.
\end{aligned}
\end{equation}

Surface distance metrics are a set of correlated measures of the distance between the surfaces of a reference and predicted lesion.

Let $S(A)$ denote the set of surface voxels of $A$. The shortest distance of an arbitrary voxel $v$ to $S(A)$ is defined as:
\begin{equation}
\begin{aligned}
d(v,S(A))=\mathop{min}\limits_{s_A\in S(A}(\left \| v-s_A \right\|),
\end{aligned}
\end{equation}
where $\left \| \bullet \right\|$ denotes the Euclidean distance.

\textbf{\emph{ASD:}} it is defined as:
\begin{equation}
\begin{aligned}
ASD(A,B)=\frac{1}{\left|S(A)\right|+\left|S(B)\right|}(\sum_{s_A\in S(A)}{d(s_A,S(B))} \\
+\sum_{s_B\in S(B)}{ d(s_B,S(A))}).
\end{aligned}
\end{equation}

\textbf{\emph{MSD:}} it is also known as the Symmetric Hausdorff Distance, and is similar to $ASD$ except that the maximum distance that is taken instead of the average:
\begin{equation}
\begin{aligned}
MSD(A,B)=max\{\mathop{max}\limits_{s_A\in S(A)}{d(s_A,S(B))},\\
\mathop{max}\limits_{s_B\in S(B)}{d(s_B,S(A))}\}.
\end{aligned}
\end{equation}

\subsection{Challenges and Future Scope}
It has been proved that fully automated segmentation of medical images based on deep neural networks is very valuable. By reviewing the progress of deep learning in medical image segmentation, we have identified potential difficulties. Researchers successfully employed a variety of means to improve the accuracy of medical image segmentation. Whereas, only the improvement of accuracy cannot account for the performance of algorithms, especially in the field of medical image analysis, where problems of class imbalance, noise interference problems and serious consequences of missed tests must be considered. In the following subsections, we will analyze potential future research directions for medical image segmentation.

\subsubsection{Design of Network Architecture}
In studies of medical image segmentation, the innovation of network structure design is most popular, as the improvement of network structure design shows clear effect and it is easily transferred to other tasks. Through reviewing classical models in recent years, we find that the basic framework of encoder-decoder U-shaped networks with long and short skipped connections has been widely used for medical image segmentation. The residual network (ResNet) and the densely connected network (DenseNet) have demonstrated the effect of deepening network depth and the effectiveness of residual structure on gradient propagation, respectively. Skip connections in deep networks can facilitate gradient propagation and thus reduce the risk of gradient dispersion leading to improved segmentation performance. Furthermore, the optimization of skipped connections will allow the model to extract richer features.

In addition, the design of the network module is worth exploring. Recently, spatial pyramid modules have been widely used in the field of semantic segmentation. The atrous convolution with fewer parameters allows for wider receptive fields, and the feature pyramid allows for features with different scales to be acquired. The development of spatial channel attention modules makes the process of neural network feature extraction more targeted, so the design of task-specific feature extraction network modules is also well worth investigating.

The manual design of model structures requires rich experiences, so it is inevitable that NAS will gradually replace the manual design. However, it is difficult to search directly a large network due to memory and GPU limitations. Therefore, the future trend should be the combination of manual design and the use of NAS technology. First, a backbone network is designed manually, and then small network modules are searched by NAS before training.

The design of different convolution operations is also a meaningful research direction, such as atrous convolution, deformable convolution, deep separable convolution, etc. Although these convolutions are all excellent for improving performance of models, they still belong to traditional convolutional categories. As a convolutional method of processing non-Euclidean data, the graph convolution goes beyond the traditional convolution and is valuable for medical data because the graph structure is more efficient and has a strong semantic feature encoding capability.

\subsubsection{Design of Loss Function}
In many medical image segmentation tasks, there are often only one or two targets in an image, and the pixel ratio of targets is sometimes small, which makes network training difficult. For this problem, it is easier to focus on smaller targets by changing loss functions than to change the network structure. However, the design of loss functions is highly task-specific, so we need to analyze carefully task requirement, and then design reasonable and available loss functions.

In specific tasks of medical image segmentation, the use of classical cross-entropy loss functions combined with a specific regularization term or a specific loss function has become a popular trend. In addition, the use of domain knowledge or a priori knowledge as regular terms or the design of specific loss functions can yield better task-specific segmentation results for medical images. Another avenue is an automatic loss function (or regularization term) search based on NAS techniques.

\subsubsection{Transfer learning}
Medical imaging is usually accompanied by severe noise interference. Moreover, the data annotation of medical images is often more expensive than natural images. Therefore, medical image segmentation based on pre-trained deep learning models on natural images is a worthy direction for future research.

In addition, transfer learning is an important way to achieve weakly supervised medical image segmentation. In fact, transfer learning is the use of existing knowledge to learn new knowledge, and it focuses on finding similarities between existing knowledge and new knowledge. Since most data or tasks are correlated, transfer learning allows us to share the model parameters (or knowledge learned by the model) with the new model in a way that speeds up the efficiency of model learning. Thus, transfer learning can solve the problem of insufficient labelling data.

\subsubsection{Interactive Segmentation}
Although deep learning has achieved good results in many image segmentation tasks, the vast majority of related works have been with automatic segmentation methods. Many cases still require interactive segmentation methods, such as the annotation of radiotherapy targets, or when user correction is required because the automatic segmentation results are not good enough. In addition, training deep learning models often requires a large number of labeled images as the training datasets that can be done more efficiently with an interactive segmentation tool.

Due to the superior performance of deep learning, the interactive image segmentation~\cite{zhang2018translating} based on deep learning can reduce the number of user interactions and the user time that shows broader application prospect.

\subsubsection{Graph Convolutional Neural Network}
In general, convolution-based deep neural networks with translation invariance, rotation invariance, scale invariance, shared convolution kernels and fast automatic feature extraction have yielded remarkable results in the field of medical images. However, convolutional neural networks also have many limitations: they rely heavily on geometric priors and it is difficult to capture the intrinsic relationships between different objects using extracted local features, etc. GNN provides a powerful and intuitive modelling approach~\cite{zhang2021affinity} to the problem of modelling non-Euclidean spaces. Taking the studied objects as nodes and the correlation or similarity between objects as edges, GNN is able to integrate non-Euclidean data and extract invisible relationships between objects by exploiting their intrinsic relationships, and it has been widely used in brain segmentation~\cite{yang2019classification}, vessel segmentation~\cite{shin2019deep}, prostate segmentation~\cite{tian2020graph}, coronary artery segmentation~\cite{yang2020cpr}, etc.

\subsubsection{Medical Transformer}
In recent years, deep neural networks based on U-shaped structures and skip connection have been widely used in various medical imaging tasks. However, in despite of the fact of achieving excellent performance by CNNs, it is unable to learn global and long-range semantic information interactions well due to the limitations of convolutional operations. Recently, transformer-based architectures have become very popular that replaces the convolutional operator and use self-attention modules to compose entire encoder-decoder structures that can encode long-range dependencies. It has been a great success in the field of natural language processing.

Dosovitskiy et al.~\cite{dosovitskiy2020image} proposed Vision Transformer (ViT) that is able to classify images directly using the Transformer. Recently, a large number of researches~\cite{chen2021transunet}~\cite{gao2021utnet}~\cite{valanarasu2021medical}~\cite{cao2021swin} have applied the transformer to medical image segmentation. CNNs have a comparative advantage in extracting the underlying features. These low-level features form the key points, lines, and some basic image structures at the patch level. However, when we detect these basic visual elements, the higher-level visual semantic information is often more concerned with how these elements relate to each other to form an object, and how the spatial location of objects relates to each other to form the scene. At present, the transformer is more natural and effective in dealing with the relationships between these elements. However, if all the convolutional operators in CV tasks are replaced by Transformer, it may suffer from many problems, such as high computational cost and memory usage. From existing researches, the combination of Transformer and CNNs may lead to better results.

\section*{Acknowledgment}

This work was supported in part by Natural Science Basic Research Program of Shaanxi (Program No. 2021JC-47), in part by the National Natural Science Foundation of China under Grant 61871259, Grant 61861024, National Natural Science Foundation of China-Royal Society: Grant 61811530325 (IECnNSFCn170396, Royal Society, U.K.), in part by Key Research and Development Program of Shaanxi (Program No. 2021ZDLGY08-07), and in part by Shaanxi Joint Laboratory of Artificial Intelligence (Program No. 2020SS-03).
\bibliographystyle{IEEEtran}

\end{document}